\documentclass[twocolumn]{aastex63}
\usepackage{amssymb}
 \usepackage{amsmath}

\usepackage{graphicx, subfigure}
\usepackage{soul}

\submitjournal{ApJ}

\shorttitle{Spectral signatures of PeVatrons}
\shortauthors{Celli et al.}
\graphicspath{{./}{figure/}}

\begin{document}

\title{SPECTRAL SIGNATURES OF PEVATRONS}

\correspondingauthor{Silvia Celli}
\email{silvia.celli@roma1.infn.it}

\author[0000-0002-7592-0851]{Silvia Celli}
\affiliation{Dipartimento di Fisica dell'Universit\`a La Sapienza, P.~le Aldo Moro 2, 00185 Rome, Italy}
\affiliation{Istituto Nazionale di Fisica Nucleare, Sezione di Roma, P.~le Aldo Moro 2, 00185, Rome, Italy}

\author{Felix Aharonian}
\affiliation{Dublin Institute of Advanced Studies, 10 Burlington Road, Dublin 4, Ireland}
\affiliation{Max-Planck-Institut f\"ur Kernphysik, Saupfercheckweg 1, D-69117 Heidelberg, Germany}

\author{Stefano Gabici}
\affiliation{Universit\'e de Paris, CNRS, Astroparticule et Cosmologie,  F-75006 Paris, France}

\begin{abstract}

We analyze the energy distributions of final (stable) products  - gamma rays, neutrinos, and electrons - produced in inelastic proton-proton collisions in the PeV energy regime.  We also calculate the energy spectrum of synchrotron radiation from secondary electrons, assuming that these are promptly cooled in the surrounding magnetic field. We show that the synchrotron radiation has an energy distribution much shallower than that of primary protons, and hence we suggest to take advantage of such a feature in the spectral analysis of the highest energy (cut-off) emission region from particle accelerators. For a broad range of energy distributions in the parent protons, we propose simple analytical presentations for the spectra of secondaries in the cut-off region. These results can be used in the interpretation of high-energy radiation from PeVatrons - cosmic-ray factories accelerating protons to energies up to 1~PeV. 
\end{abstract}


\section{Introduction}
\label{sec:Intro}
The gamma-ray observations of recent years have unveiled thousands of accelerators of relativistic particles linked to almost all known non-thermal Galactic and extra-galactic source populations; see, in particular,  the recent compilations based on  the H.E.S.S. Galactic Plane Survey \citep{hessGPS}, the Fermi-LAT Fourth Source Catalog \citep{fermi2020}, and the Third HAWC Catalog of Very-High-Energy Gamma-Ray Sources \citep{3hawc}. The broad range of implications of these discoveries concerns several areas related, in particular, to the origin of Cosmic Rays (CRs), the physics and astrophysics of relativistic outflows (e.g. the pulsar winds and Active Galactic Nuclei jets), the search for Dark Matter, {\it etc.}  In the context of the origin of Galactic CRs, a prime interest is represented by hadronic accelerators, especially those capable of accelerating protons and nuclei up to PeV energies, the so-called Cosmic PeVatrons. The extension of the spectrum of Galactic CRs to the so-called  {\it knee}, around $E_{\rm p,0} \sim 1$~PeV (see, e.g. \citet{gabiciReview}), is an indication for the existence of such CR factories in our Galaxy. 

Despite the discovery of tens of TeV gamma-ray emitters, including the ones associated with young Supernova Remnants (SNRs), the suspected major contributors to the Galactic CRs, we have only limited information about proton PeVatrons.  Possible candidates are, in particular,  the source(s) in the Galactic Center region \citep{HESSgc}, and a few extended regions surrounding young stellar clusters \citep{FA2018}.
Recently, some more candidates have been reported by the HAWC Collaboration \citep{3hawc}. The search for proton PeVatrons is considered as one of the priorities of the ground-based gamma-ray astronomy. The start of operation of the Large High Altitude Air Shower Observatory (LHAASO) with superior gamma-ray sensitivity above 100~TeV, and the upcoming Cherenkov Telescope Array (CTA) with excellent pointing capabilities and angular resolution, promise a breakthrough in this area. 

Galactic PeVatrons have distinct spectral signatures. The final neutral and stable products of proton-proton ($pp$) collisions, i.e. gamma rays and neutrinos, receive approximately $\sim 10\%$ of the energy of primary protons. Therefore $\geq 100$~TeV gamma rays and neutrinos carry straightforward and model-independent information about the primary PeV protons. Hence, spectrometry well beyond the cut-off region $E_{\gamma, 0}\sim 0.1 E_{\rm p,0} \sim 100$~TeV is of paramount importance for the identification of the acceleration mechanism and the conditions operating in the acceleration region.  Spectral measurements of PeVatrons above $100$~TeV can be realised by future ground-based gamma-ray detectors (both with CTA and LHAASO). This should be the case of SNRs characterised by hard power-law energy distribution and a relatively slow (e.g. exponential) cut-off.  For steeper acceleration spectra,  a proper spectrometry would be problematic even for future generation instruments. Besides, if the PeVatrons are located in enhanced far-infrared radiation regions, e.g. in the Galactic Center region, one may expect significant distortion of the initial energy spectrum of $\geq$100~TeV gamma rays due to the $\gamma-\gamma$ absorption in pair production process \citep{celli2017}. At these conditions, neutrinos would act as unique messengers as they do not suffer such absorption. Still, even in the most optimistic scenarios, neutrino spectroscopy appears challenging for the current km$^3$-scale neutrino detectors \citep{ambrogi2018}. Meanwhile, complementary information about PeVatrons is carried by synchrotron photons emitted by secondary electrons in the surrounding magnetic fields \citep{FSaasFe}. 

The calculations of characteristics of gamma rays, neutrinos and electrons as the final products of proton-proton interactions include integrations over inclusive differential cross-sections of short-lived secondaries ($\pi$ and  K-mesons, {\it etc.}) and the kinematic relation of their decays. In the case of broad energy distributions, one can avoid extensive integrations, by using a simple $\delta$-functional approach instead. The latter is a rather good approximation, provided that the distribution of the incident protons does not contain sharp spectral features \citep{AA2000}. Otherwise, this approximation leads to wrong results. In particular, in the case of the power-law distribution of protons with an exponential cut-off, the $\delta$-functional approach predicts the gamma-ray spectrum to closely mimic the parent proton spectrum, with a shift towards lower energies. However, the detailed numerical calculations in the cut-off region show a significantly shallower gamma-ray distribution \citep{kelner2006}. 

The aim of this paper is a detailed study of the energy spectra of gamma rays, neutrinos and synchrotron radiation of the secondary electrons emerging from $pp$ collisions, 
with a focus on the spectral signatures of secondaries linked to the highest energy protons from the cut-off region. The latter contains crucial information about the conditions operating inside the accelerator, concerning, in particular, the energy-dependent acceleration rate, the efficiency of the confinement and escape of the relativistic particles from the accelerator, {\it etc}. Representing the energy distributions of accelerated protons in the form of power-law with a generalized exponential cut-off, $\propto E^{-\alpha} \exp[-(E/E_0)^\beta$, we offer simple analytical presentations of $\alpha, \beta$ and $E_0$, that can conveniently be used in the data reduction and interpretation of observations. 

The paper is structured as it follows: in Sec.~\ref{sec:methods} we introduce the energy distribution of the projectile protons, discussing its parametric representation and normalization. We outline the method for calculating the spectra of secondary particles in Sec.~\ref{sec:methods2}, providing parametrizations of the spectral parameters in Sec.~\ref{sec:param}. In Sec.~\ref{sec:maxwell} we consider the case of Maxwellian-like energy distribution of primary protons. Results are summarized and discussed in Sec.~\ref{sec:disc}. The computation of secondary particle spectra is complemented with Appendix~\ref{sec:app}, where a collection of the relevant equations is given.

\section{Energy distribution of protons}
\label{sec:methods}
In $pp$ collisions, for the given energy distribution of parent protons, the energy spectra of secondaries are determined by inclusive cross-sections. In this work, unless otherwise stated, we will represent the energy distribution of relativistic protons in the following form: 
\begin{equation}
\label{eq:protonsS}
J_{\rm p} \equiv \frac{dN_{\rm p}}{dE_{\rm p}dV} = K_{\rm p} E_{\rm p}^{-\alpha_{\rm p}} \exp\left[-\left( \frac{E_{\rm p}}{E_{0,{\rm p}}}\right)^{\beta_{\rm p}}\right] \ .
\end{equation}
It consists of the power-law part with slope $\alpha_{\rm p}$, and of the cut-off at energy $E_{0,{\rm p}}$, 
in a general exponential form defined by the index $\beta_{\rm p}$.
The normalization constant $K_{\rm p}$ is determined  by the condition of the energy density above 100~GeV to be 
\begin{equation}
\label{eq:pdis}
w_{\rm p} = \int_{100~\rm{GeV}}^{\infty} E_{\rm p} J_{\rm p} (E_{\rm p}) dE_{\rm p} = 1~\rm{erg}~\rm{cm}^{-3} \ .
\end{equation}
The $100$~GeV energy in protons translates into $\sim 10$~GeV energy  gamma rays and neutrinos, significantly above the so-called pion-bump region determined by the kinematics of pion decays. 

The energy distribution of protons given by Eq.~(\ref{eq:protonsS}) includes three parameters, $\alpha_{\rm p}$, $\beta_{\rm p}$ and $E_{0,{\rm p}}$. With different combinations of these parameters, one can describe a broad range of distributions of protons accelerated in different astrophysical environments. For example, the case of Diffusive Shock Acceleration (DSA) provides $\alpha_{\rm p} \sim 2$ in the test-particle limit. But in more realistic scenarios  $\alpha_{\rm p}$ could be larger or smaller 2, depending on the conditions characterizing the acceleration region. E.g. the particle feedback on the shock itself tends to produce harder spectra, with $\alpha_{\rm p} \sim 1.5$ \citep{malkov2001}. For acceleration mechanisms different from DSA, e.g. for some versions of stochastic acceleration or magnetic reconnection \citep{lazarian2015}, the distribution of accelerated particles could be rather narrow, e.g. of Maxwellian type. In this case, the distribution can be described by a small value of  $\alpha_{\rm p} \leq 0$. 

The parameters $E_{0,{\rm p}}$ and  $\beta_{\rm p}$  characterize the efficiency of acceleration at highest energies.  In the case of PeVatrons, the cut-off energy should exceed (by definition) 0.1~PeV. In accelerators responsible for protons well above the \textquoteleft\textquoteleft knee\textquoteright\textquoteright $\,$ in the spectrum of Galactic CRs,  $E_{0,{\rm p}}$ should be as large as 10 PeV. Concerning the parameter $\beta_{\rm p}$, it is often fixed to $\beta_{\rm p}=1$. Such a shape is predicted, in particular, by the standard DSA theory, when the diffusion is close to the Bohm regime. However, in general, depending on the conditions in the acceleration region,
$\beta_{\rm p}$ can deviate from 1. This could happen, for example,  in the case when the particle diffusion at the highest energies operates in a regime different from the Bohm diffusion one, or when the losses, e.g. due to interactions or escape from the acceleration zone, become non-negligible. The shape of the accelerated spectrum in the cut-off region not only depends on the acceleration mechanism but it is also very sensitive to the conditions in the accelerator zone. Nevertheless, its representation in the form of generalized exponential cut-off with two parameters,  $E_{0,{\rm p}}$ and $\beta_{\rm p}$, is not only a convenient mathematical presentation but can describe a broad range of acceleration scenarios. 

\section{Energy distributions of secondary products}
\label{sec:methods2}
Proton-proton collisions proceed at a rate dictated by the interaction cross-section and the density $n$ of target particles. Concerning the total inelastic cross-section of proton-proton collisions, we will adopt the latest available parametrization \citep{kafexhiu2014}, also accounting fo LHC measurements: 
\begin{equation}
\label{eq:sigmapp}
\begin{split}
\sigma_\textrm{inel} (T_{\rm p}) = & \left[30.7 - 0.96 \log \left(\frac{T_{\rm p}}{T_{\rm p}^{\rm th}} \right) + 0.18 \log^2 \left(\frac{T_{\rm p}}{T_{\rm p}^{\rm th}} \right) \right] \\
& \times \left[ 1- \left(\frac{T_{\rm p}^{\rm th}}{T_{\rm p}} \right)^{1.9} \right]^3 \, \textrm{mb} \ ,
\end{split}
\end{equation}
where $T_{\rm p}$ is the kinetic energy of the incident proton, and
$T_{\rm p}^{\rm th}\simeq 0.2797$~GeV is the threshold for the neutral pion production. Note that at TeV energies, the cross-section given by Eq.~\eqref{eq:sigmapp} is larger, by $\approx 20 \%$, compared to the parametrization of the total cross-section adopted in \citet{kelner2006}. 

\subsection{Gamma rays}
\label{subsec:gamma}
High-energy photons are mainly generated in the decay of the neutral pions produced at $pp$ collisions, with an order of magnitude smaller contribution arising from the decay of $\eta$ mesons. Following \citet{kelner2006}, the gamma-ray emissivity can be written as
\begin{equation}
\label{eq:phigamma}
\epsilon_\gamma(E_\gamma) = c n \int_{0}^{1} \frac{dx}{x} \sigma_\textrm{inel} \left( \frac{E_\gamma}{x} \right) J_{\rm p} \left( \frac{E_\gamma}{x} \right) F_\gamma \left(x, \frac{E_\gamma}{x} \right) \ ,
\end{equation}
where $c$ is the speed of light in vacuum, and $x=E_\gamma/E_{\rm p}$. $J_{\rm p}(E_{\rm p})$ is the energy distribution of protons given  by  Eq.~\eqref{eq:protonsS}, and $F_\gamma (x, E_{\rm p})$ is the so-called kernel function. Below we use the analytical parametrizations of these functions given by  Eqs.~(58)-(61) of \citet{kelner2006}.  For the convenience of the reader, all these equations are compiled in Appendix~\ref{sec:app}. 

\subsection{Neutrinos}
\label{subsec:nu}
The muon and electron neutrinos are produced at the decays of  charged pions and muons, respectively. The emissivity of muon neutrinos and antineutrinos in the full decay chain, i.e. for both the channels of the pion (1) and muon (2) decays, reads
\begin{equation}
\label{eq:phinu}
\begin{split}
\epsilon_{\nu_\mu}(E_\nu) =  & c n \left[ \int_{0}^{r_\pi} \frac{dx}{x} \sigma_\textrm{inel} \left( \frac{E_\nu}{x} \right) J_p \left( \frac{E_\nu}{x} \right) F^{(1)}_{\nu_\mu} \left(x, \frac{E_\nu}{x} \right) + \right. \\
& \left. + \int_{0}^{1} \frac{dx}{x} \sigma_\textrm{inel} \left( \frac{E_\nu}{x} \right) J_p \left( \frac{E_\nu}{x} \right) F^{(2)}_{\nu_\mu} \left(x, \frac{E_\nu}{x} \right) \right] \ , 
\end{split}
\end{equation}
where $x=E_\nu/E_{\rm p}$ and $r_\pi=0.427$. For the kernel functions $F^{(1)}_{\nu_\mu} (x, E_{\rm p})$ and $F^{(2)}_{\nu_\mu} (x, E_p)$ we adopt Eqs.~(66)-(69) and Eqs.~(62)-(65) of \citet{kelner2006} (see also  Appendix~\ref{sec:app}). 

The emissivity of the electron neutrinos, which arise from the muon decays, is 
\begin{equation}
\label{eq:phie}
\epsilon_{\nu_{\rm e}} (E_{\nu_{\rm e}}) =  c n \int_{0}^{1} \frac{dx}{x} \sigma_\textrm{inel} \left( \frac{E_\nu}{x} \right) J_p \left( \frac{E_\nu}{x} \right) F_{\nu_{\rm e}} \left(x, \frac{E_\nu}{x} \right) \ ,
\end{equation}
where $F_{\nu_{\rm e}}  \approx F^{(2)}_{\nu_\mu}$, with an accuracy better than 5\% \citep{kelner2006}.
The kaon decay chain proceeds similarly to the pion decays.
While at low energies the neutrino contribution arising from this addition channel to the total neutrino flux is about  10\%, at highest energies it is significantly smaller \citep{kelner2006}. Therefore, below  we ignore the neutrino production channels related to the kaon decays.  

\subsection{Secondary electrons}
\label{subsec:e}
{The spectra of electrons and positrons emerge as the final products of the chain of decays with the involvement of charged pions and muons.  Hereafter we describe both electrons and positrons with the same term  \textquoteleft\textquoteleft electrons\textquoteright\textquoteright.  Their spectra  closely resemble that of electron neutrinos (antineutrinos), $\epsilon_{\rm e}(E) \approx \epsilon_{\nu_{\rm e}}(E)$ \citep{kelner2006}. 

Once the electron production rate is derived, we assume that the electron energy distribution is established, i.e. it has achieved a steady-state condition through the synchrotron cooling. This implies that the synchrotron cooling time of electrons of energy $E_{\rm e}$ in the magnetic field $B_0$,
\begin{equation}
\label{eq:tausynch}
\tau_\textrm{sy}(E_{\rm e}) \simeq 12.5 \left(\frac{E_{\rm e}}{1~\textrm{TeV}} \right)^{-1} \left(\frac{B_0}{1~\textrm{m}\textrm{G}} \right)^{-2} \, \textrm{yr} \  ,
\end{equation}
does not exceed the characteristic dynamical timescales, in particular the energy loss time of the parent protons and the age of the accelerator. Using  the relation between the average energy of the synchrotron photon $\bar{E}_{\rm sy}$ and the energy of the parent electron
\begin{equation}
\label{eq:esy}
\bar{E}_{\rm sy} \simeq 0.02 \left( \frac{B_0}{\textrm{mG}} \right) \left( \frac{E_{\rm e}}{{\rm TeV}} \right)^2 \, \textrm{keV} \ ,
\end{equation}  
the cooling time of electrons can be expressed as a function of $\bar{E}_{\rm sy}$ and $B_0$ as:
\begin{equation}
\label{eq:taucool}
\tau_\textrm{sy}(\bar{E}_{\rm sy}) \simeq 1.7 \left(\frac{\bar{E}_{\rm sy}}{1~\textrm{keV}} \right)^{-1/2} \left(\frac{B_0}{1~\textrm{m}\textrm{G}} \right)^{-3/2} \, \textrm{yr} \ .
\end{equation}  

The average energy loss time of protons due to the inelastic $pp$ interactions is almost energy-independent  
$\tau_{\rm pp} \simeq 1.7 \times 10^7 ( n/{\rm cm}^{-3} )^{-1}$~yr, 
where $n$ is the number density of the ambient gas.  Thus, even in very dense environments like  giant molecular clouds with density exceeding 
$10^4 \ \rm cm^{-3}$, the characteristic cooling time of protons is longer than hundreds of years.  Therefore, the age of the accelerator is a more critical issue. For example, although young supernova remnants of age of 
$10^3$~yr formally can operate as PeVatrons, the careful theoretical treatments give a preference 
to the acceleration of PeV protons at much earlier epochs of  SNR evolution, $\leq 10$ years (see, e.g. \citealt{bell2013}).  Yet, the cooling time of electrons responsible for the production of synchrotron X-ray emission is shorter, provided that the magnetic field is not much weaker than 1~mG, the latter being a 
key condition for the realization of acceleration of protons to PeV energies. 
As it follows from Eq.~\eqref{eq:tausynch},  the electron cooling time 
decreases linearly with the energy, therefore the steady state is established only at energies
above $E^\star$ which is determined 
from the condition $T_0=\tau_{\rm sy}(E^\star)$. Below, for sake of simplicity, we assume that all electrons are cooled down. Note that this formal assumption is not critical for our main objective  - the study of hard X-ray signatures of PeVatrons represented by the synchrotron radiation of secondary electrons linked to the highest energy protons from the cut-off region.

\subsection{Synchrotron radiation of secondary electrons}
\label{subsec:ph}

The steady-state electron energy distribution, due to complete synchrotron cooling, is obtained from the $pp$ emissivity as
\begin{equation}
\label{eq:SSelectrons}
J_{\rm e}(E_{\rm e}) = \frac{\tau_\textrm{sy}(E_{\rm e})}{E_{\rm e}} \int_{E_{\rm e}}^{\infty} \epsilon_{\rm e}(E) dE \ .
\end{equation}
Then, the synchrotron emissivity from such electrons radiating in an isotropic magnetic field of strength $B_0$ is
\begin{equation}
\label{eq:synch}
\epsilon_{\rm sy}(E)=\frac{\sqrt{3}}{2\pi} \frac{e^3 B_0}{m_{\rm e} c^2} \frac{1}{\hbar E} \int_0^{\infty} J_{\rm e}(E_{\rm e}) R \left(\frac{E}{E_{\rm c}(E_{\rm e})}\right ) dE_{\rm e} \ ,
\end{equation}
where $e$ is the electron's charge, $m_{\rm e}$ its mass, $\hbar$ is the reduced Planck's constant, and $E_{\rm c}(E_{\rm e})$ is the characteristic energy of synchrotron photons emitted by an electron of energy $E_{\rm e}$, such that $\bar{E}_{\rm sy} \simeq 0.29 E_{\rm c}$.

Below we normalize the magnetic field to  $B_0=1$~mG. 
Introducing  $x=E/E_{\rm c}$, the function $R(x)$ for the field of constant strength $B_0$ is \citep{zirakashvili2010}
\begin{equation}
\label{eq:Buniform}
R(x)=\frac{1.81 e^{-x}}{\sqrt{x^{-2/3}+(3.62/\pi)^2}} \ .
\end{equation}

In  turbulent environments, the magnetic field is characterized by a distribution not only over directions but also over the strength. Thus, one should include the B-field strength probability distribution in  calculations of the synchrotron spectrum. Here we assume that the magnetic field has  Gaussian distribution:
\begin{equation}
P(B)= \left(\frac{6}{\pi} \right)^{1/2} \frac{3B^2}{B^3_0} \exp \left[- \frac{3B^2}{B_0^2} \right] \ , 
\end{equation}
with the average intensity $\langle B^2 \rangle = B_0^2$. 
For this distribution, the synchrotron emissivity reads \citep{derishev2019}
\begin{equation}
\label{eq:Bturbulent}
R(x^\prime)=\frac{\alpha}{3\gamma^2_e} \left( 1+ \frac{1}{{x^\prime}^{2/3}} \right) e^{-2{x^\prime}^{2/3}} \ ,
\end{equation}
where $\alpha \simeq 1/137$ is the fine structure constant and $x^\prime=9x/8$. 

A caveat linked to the discussion of the previous section, is that in the steady-state solution of the electron transport, that we are here considering through Eq.~\eqref{eq:SSelectrons}, no spectral break appears, both in the distribution of electrons and in the synchrotron radiation. Typically, this assumption is  justified for the high-energy part of the electron spectrum, where the cooling time does not exceed the characteristic dynamical scale of the system. However, at low energies, the electrons might not have enough time to cool completely. Therefore, their distribution would maintain, at energies below the cooling break, the shape of the injection spectrum. 
Correspondingly, the synchrotron radiation would also have a hard spectrum below the break.

\section{Parametrizations}
\label{sec:param}
The spectra of secondaries in the cut-off region contain direct information about the proton spectrum at the highest energies, thus they can shed light on the  acceleration processes and physical conditions in the acceleration sites. Motivated by the recent exciting discoveries of multi-TeV gamma rays from a large number objects representing different astrophysical source populations, we conduct a detailed numerical study of the spectral features of secondaries for a broad range of energy distributions of protons represented in the form of Eq.~\eqref{eq:protonsS}. For such an energy distribution, we perform calculations for the following set of parameters $\alpha_{\rm p}=[1.5,2.0,2.5]$, $\beta_{\rm p}=[0.5,1.0,1.5,2.0]$ and $E_{0,{\rm p}}=[10^5,10^6,10^7, 10^8]$~GeV. The distributions of all stable secondaries are shown  in Fig.~\ref{fig:spectra} for different combinations of $\alpha_{\rm p}$, $\beta_{\rm p}$, and $E_{0,{\rm p}}$. For calculations of the secondary products of $pp$ interactions, the density of the  ambient hydrogen gas is normalized to $n=1$~cm$^{-3}$, and the energy density of relativistic protons above 100~GeV to $w_{\rm p}=1 \rm \ erg/cm^3$. In Fig.~\ref{fig:gammaNuA}, we highlight the distributions of gamma rays, muon neutrinos and electron neutrinos for $\alpha_{\rm p}=2$, $\beta_{\rm p}=1$ and $E_{0,{\rm p}}=10^6$~GeV. Note that the spectrum of electron neutrinos coincides with the electron spectrum. We also report in Fig.~\ref{fig:gammaNuB} the ratio among the neutrino and gamma-ray spectra, in terms of differential number of particles, for both muon and electron flavors.

In Fig.~\ref{fig:synch}, we show the spectra of synchrotron radiation produced in uniform and turbulent magnetic fields, for different field strength ranging from $B_0=10^{-5}$~G to $B_0=0.1$~G. Fixing $B_0$, at low energies of synchrotron radiation, corresponding to the power-law part of the parent electron distribution, the spectra derived in uniform and turbulent distributions coincide. At higher energies of synchrotron photons, $E_{\rm sy} \geq 1 \ \rm keV$, produced by electrons from the cut-off region ($E_{\rm e} \geq 10^4 \ \rm GeV$), one can see a noticeable deviation. In the cut-off region, the radiation produced in the turbulent field is somewhat flatter than in the uniform field. The behavior of synchrotron spectra for different magnetic field intensities shows interesting features: we observe that by increasing the field strength the peak energy increases, as expected from Eq.~\eqref{eq:esy}. However, the flux at the peak is not affected by the value of the magnetic field, as we are working in the hypothesis of complete cooling of secondary electrons, namely an optically thick target where all electrons energy is converted to synchrotron photons. In turn, for an optically thin target, we should observe a shift of the flux normalization towards higher (lower) values for larger (smaller) field strengths, in addition to the shift on the peak energy. Further, we will only show the spectra of radiation formed in the Gaussian turbulent field. 

In the cut-off region, the spectrum of synchrotron radiation is significantly shallower than the spectrum of parent electrons as well as the spectra of gamma rays and neutrinos. This makes the synchrotron radiation of secondary electrons a potentially more powerful tool (than gamma rays and neutrinos) for studying the spectral features of PeVatrons at energies well beyond the cut-off, $E \geq 10 E_{0,\rm p}$.  

Because of the gradual increase of the inelastic cross-section of $pp$ interactions (see Eq.~\eqref{eq:sigmapp}), the spectra of secondary gamma rays, neutrinos and electrons are slightly harder than the spectra of parent protons:  
$\alpha_{\nu} \simeq \alpha_{\gamma} \simeq \alpha_{\rm p}-0.1$.
This can be seen in Fig.~\ref{fig:gammaNuA}; the spectra of secondaries in the energy band far both from the kinematic threshold and the high energy cut-off regions, are slightly harder than $E^{-\alpha_{\rm p}}$ spectrum of the parent protons. 
On the other hand, for the power-law proton spectrum with a cut-off represented in the form of Eq.~\eqref{eq:protonsS},  the exponential term not only results in a dramatic suppression of the fluxes of secondaries at energies above $E_{\rm s} \sim  0.1 E_{\rm 0,p}$, but it causes a gradual steepening of the spectrum before the cut-off region. The effect is especially strong 
for the parameter  $\beta_{\rm p} \leq 1$.  For example, for the proton energy distribution  with $\alpha_{\rm p}=2$ and $\beta_{\rm p}=1$, the spectrum of gamma rays contains an exponential term  which can be approximated as $\exp{[-(16 E /E_{0,\rm p})^{1/2}}]$ \citep{kelner2006}. 
The impact of this term on the gamma-ray spectrum becomes substantial (more than 10\%) already at gamma-ray energies as small as 
$E_\gamma \simeq 10^{-3} E_{0,{\rm p}}$.  At intermediate energies, this effect partly compensates the spectral hardening of the gamma-ray spectrum because of the increase of the integral $pp$ cross-section with energy. This implies that for the correct determination  of the spectrum of parent protons, namely the extraction of the parameters $\alpha_{\rm p}$, 
$\beta_{\rm p}$ and $E_{0,\rm p}$,  we need broad-band gamma-ray data, typically over 3-4 decades in energy.

As the spectra of secondaries resemble the spectrum of parent protons,
we proceed with a fitting of the spectral energy distributions of secondary particles by a generic  \textquoteleft\textquoteleft exponentially suppressed power-law\textquoteright\textquoteright $\,$ function.
Namely, in analogy with Eq.~\eqref{eq:protonsS}, we represent the spectra of secondary species by the distribution 
\begin{equation}
\label{eq:fitPLC}
J_{\rm s} (E) = K_{\rm s} E^{-\alpha_{\rm s}} \exp\left[-\left( \frac{E}{E_{0,{\rm s}}}\right)^{\beta_{\rm s}}\right] \ ,
\end{equation}
where the subscript $s$ refers to secondaries:  $s=e$ for electrons, $s=\gamma$ for gamma rays, $s=\nu_\mu$ for muon neutrinos, $s=\nu_{\rm e}$ for electron neutrinos, and $s=sy$ for the  synchrotron radiation of secondary electrons. We aim at deriving analytical relations between the three parameters characterizing the energy distributions of secondaries and of parent protons. 

The modeling of the particle spectral distributions is achieved by fitting the spectra for each particle species, with a fixed minimum energy of the fit equal to $E_{\rm min} \simeq  30$~GeV for electrons, neutrinos and gamma rays. We then determine the maximum energy of the fit by requiring a reduced $\chi^2$ ($\chi^2/{\rm n}_{\rm dof}$, where ${\rm n}_{\rm dof}$ is the number of degrees of freedom) of order unity. As a result, the energy range adopted for this multifrequency fit extends from $\sim 30$~GeV to $\sim 1.5E_{0,p}$ for gamma rays and neutrinos, while it spans from $\sim 0.01$~eV to $\sim 10^{-1}(E_{0,p}/10^6 \, \textrm{GeV})^2$~GeV for synchrotron photons (in a mG magnetic field). We note that the modeling is only marginally affected by the energy range chosen for the fitting procedure. Apparently, in order to recover the power-law part of the spectrum, one should perform the fit far enough from the cut-off region. Analogously, in order to explore the cut-off region, one should perform the fit far from the pure power-law domain. In the following, we present the results of a broadband spectral modeling. 

%
\begin{figure*}
\centering
\subfigure[]{\includegraphics[width=0.48\textwidth]{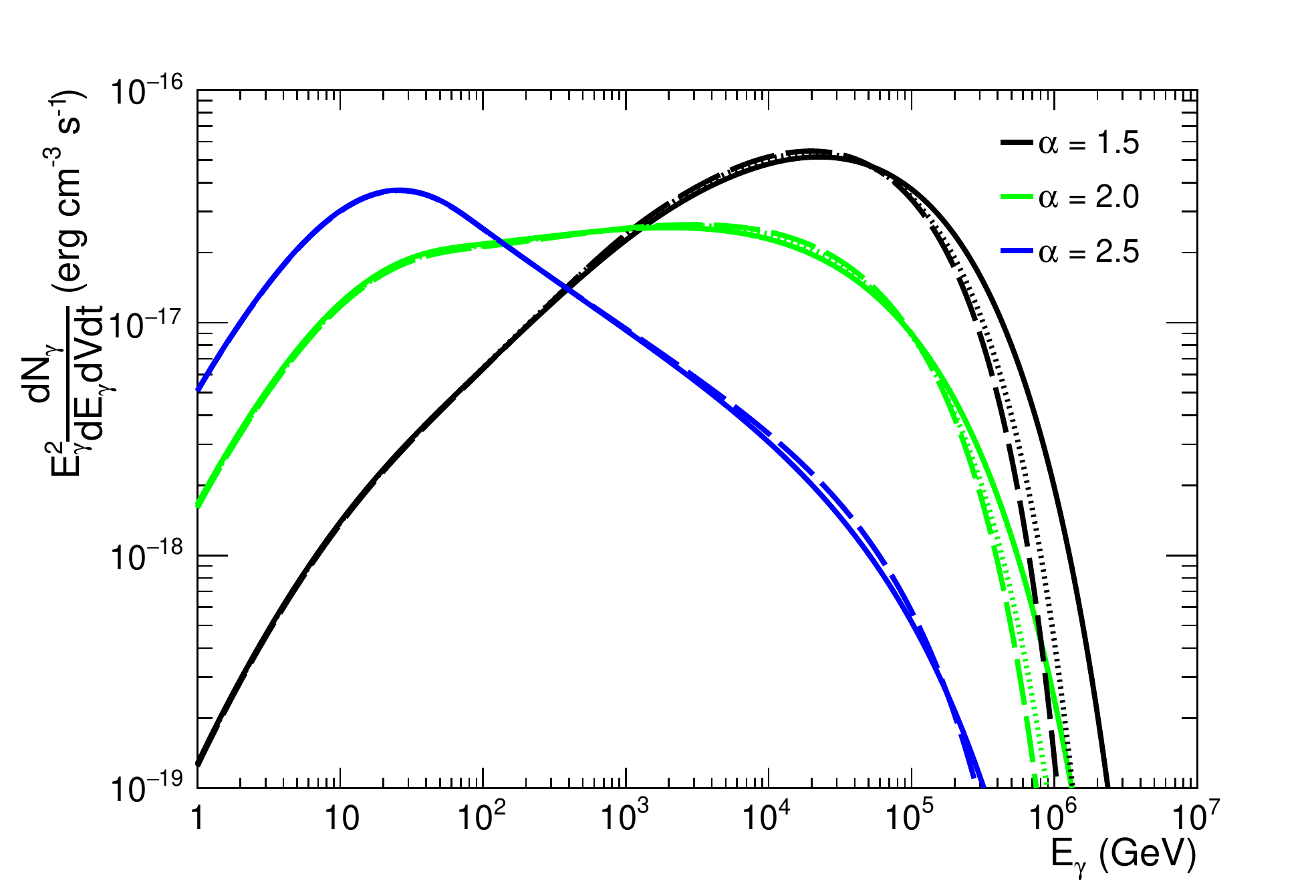}}
\subfigure[]{\includegraphics[width=0.48\textwidth]{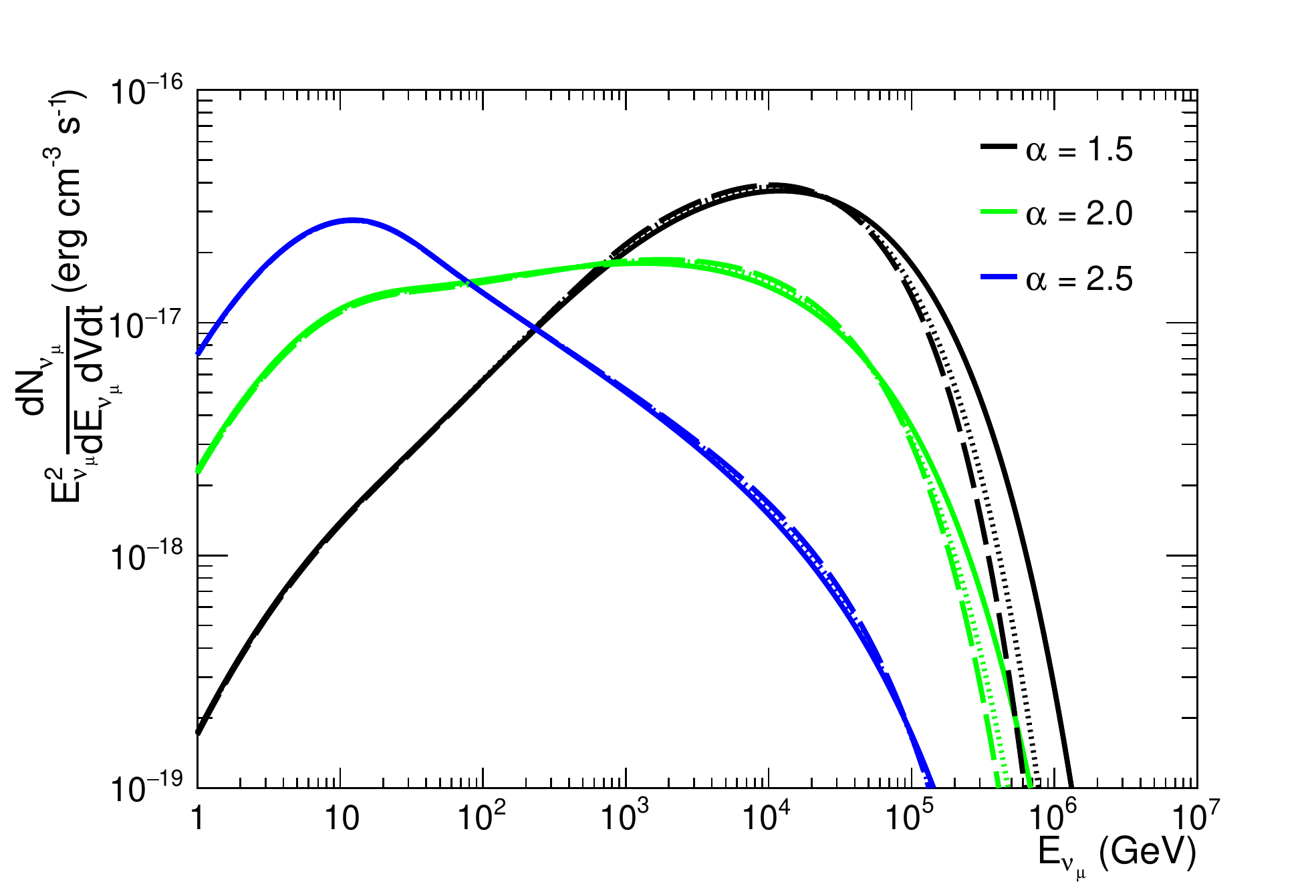}}
\subfigure[]{\includegraphics[width=0.48\textwidth]{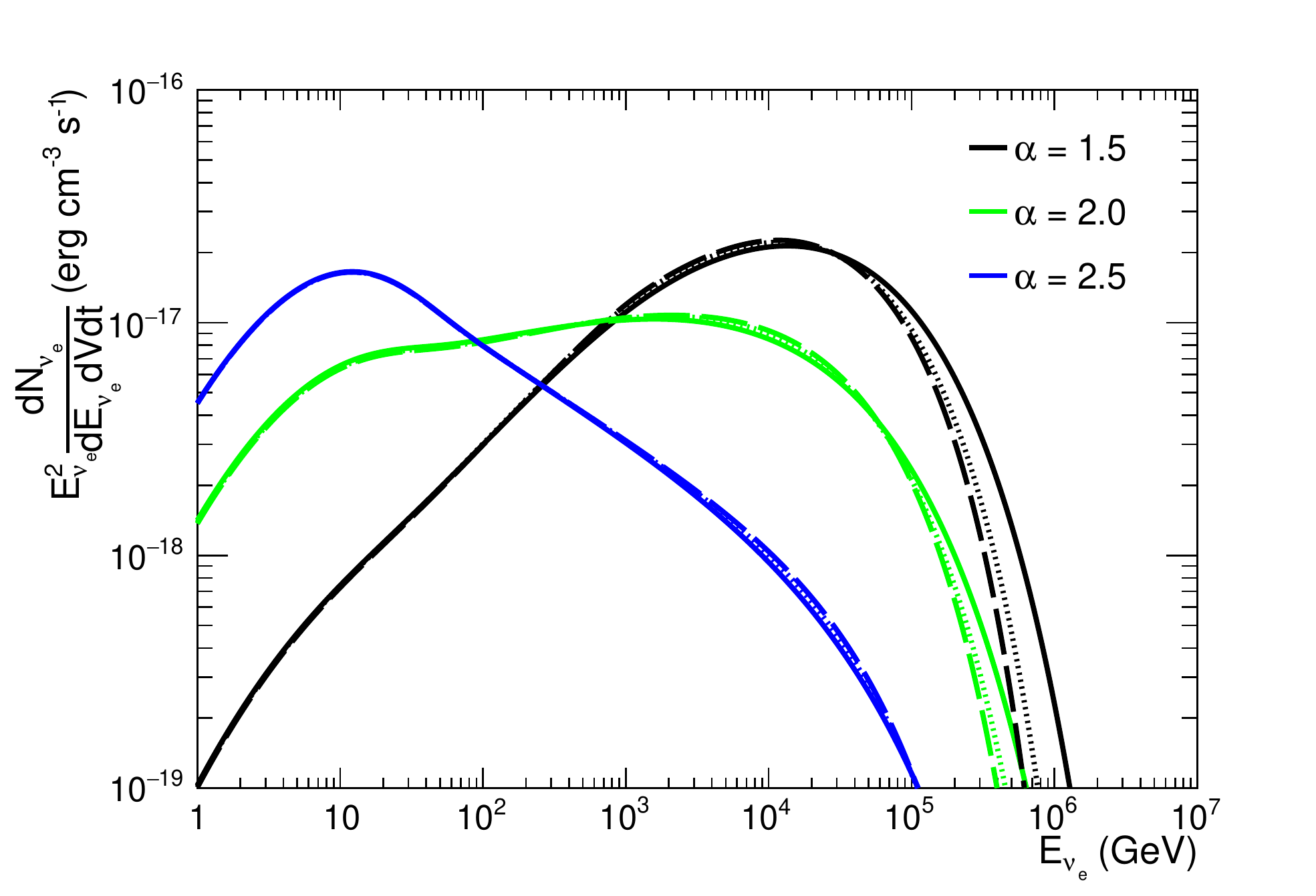}}
\subfigure[]{\includegraphics[width=0.48\textwidth]{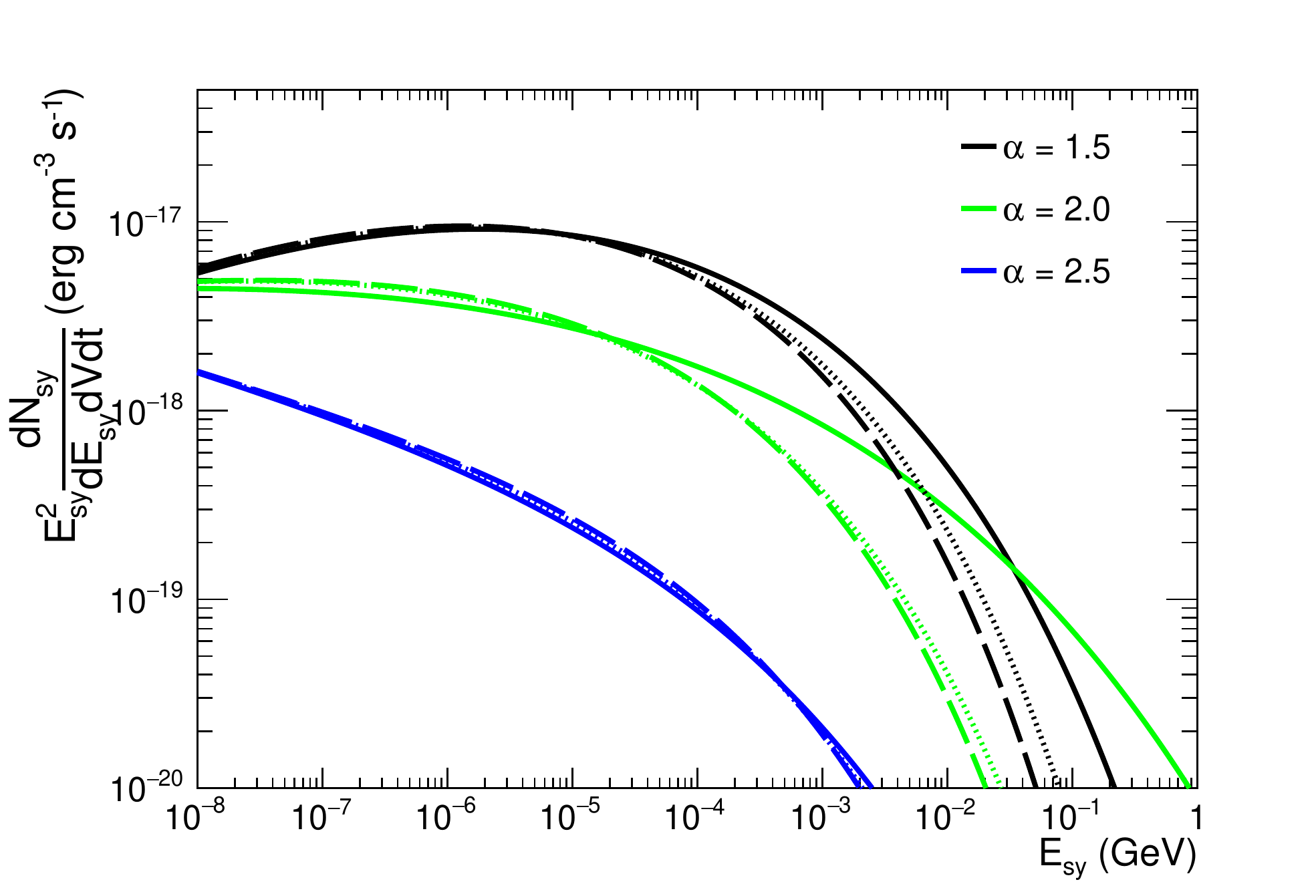}}
\caption{Emissivities of secondary products from $pp$ interactions: (a) gamma rays from $\pi^0$- and $\eta$ meson decays; (b) muon neutrinos from $\pi^\pm$ and $\mu^\pm$ decays; (c) electron neutrinos from $\mu^\pm$ decays; (d) synchrotron radiation of secondary electrons (and positrons) in a uniform magnetic field with
$B_0=1$~mG. The spectrum of parent protons 
is assumed in the form given by Eq.~\eqref{eq:protonsS}
with the following parameters: 
$E_{0,{\rm p}}=10^6$~GeV, $\alpha_{\rm p}=$1.5 (black curves), 2 (green curves), 2.5 (blue curves), and 
$\beta_{\rm p}=1.0$ (solid curves), 1.5 (dotted curves), 2 (dashed curves). The  spectrum of the parent protons is normalized to an energy density above 100~GeV of $w_{\rm p}=1 \ \rm erg/cm^3$; the number density of the ambient hydrogen gas is $n=1 \ \rm cm^{-3}$. }
\label{fig:spectra}
\end{figure*}

%
\begin{figure*}
\centering
\subfigure[\label{fig:gammaNuA}]{\includegraphics[width=0.48\textwidth]{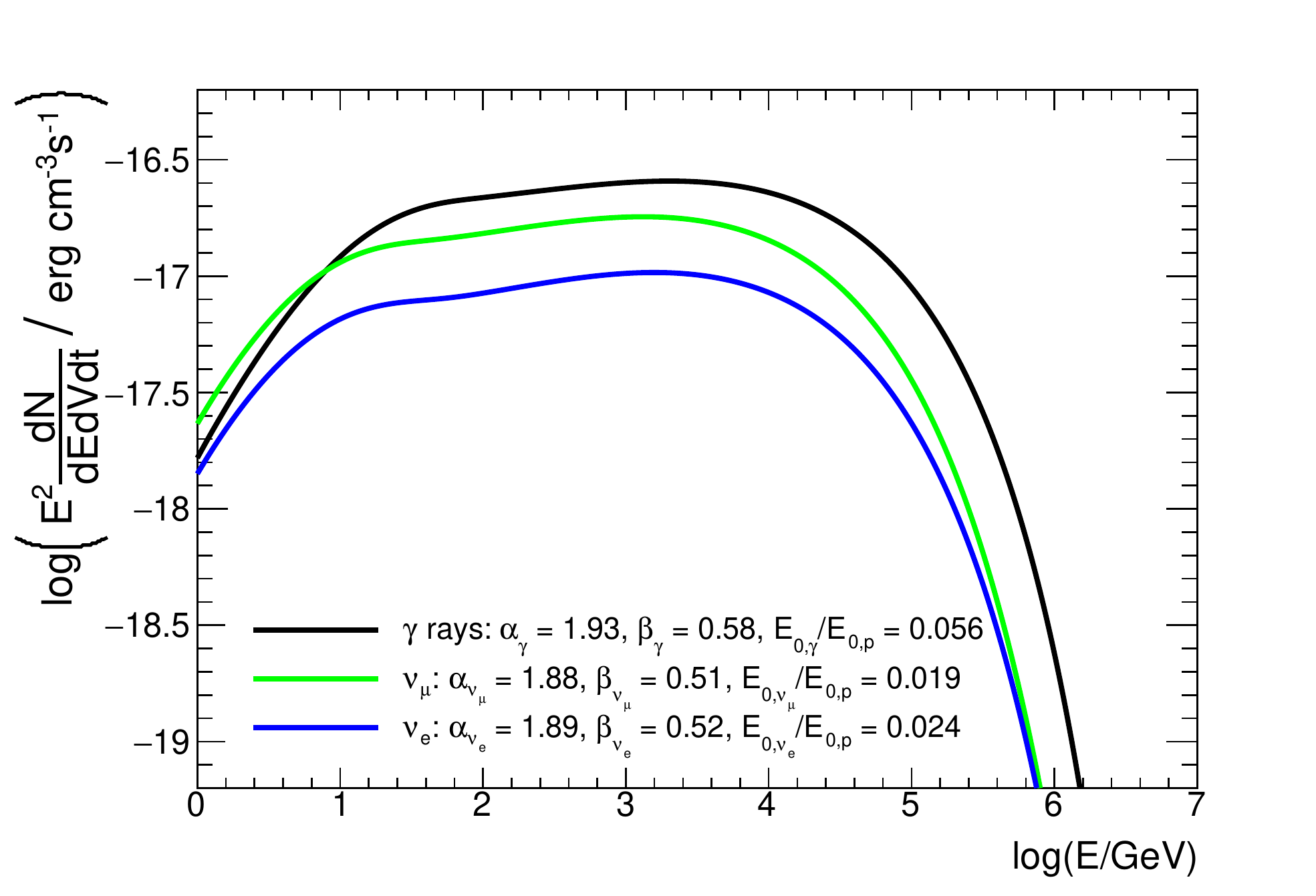}}
\subfigure[\label{fig:synch}]{\includegraphics[scale=0.43]{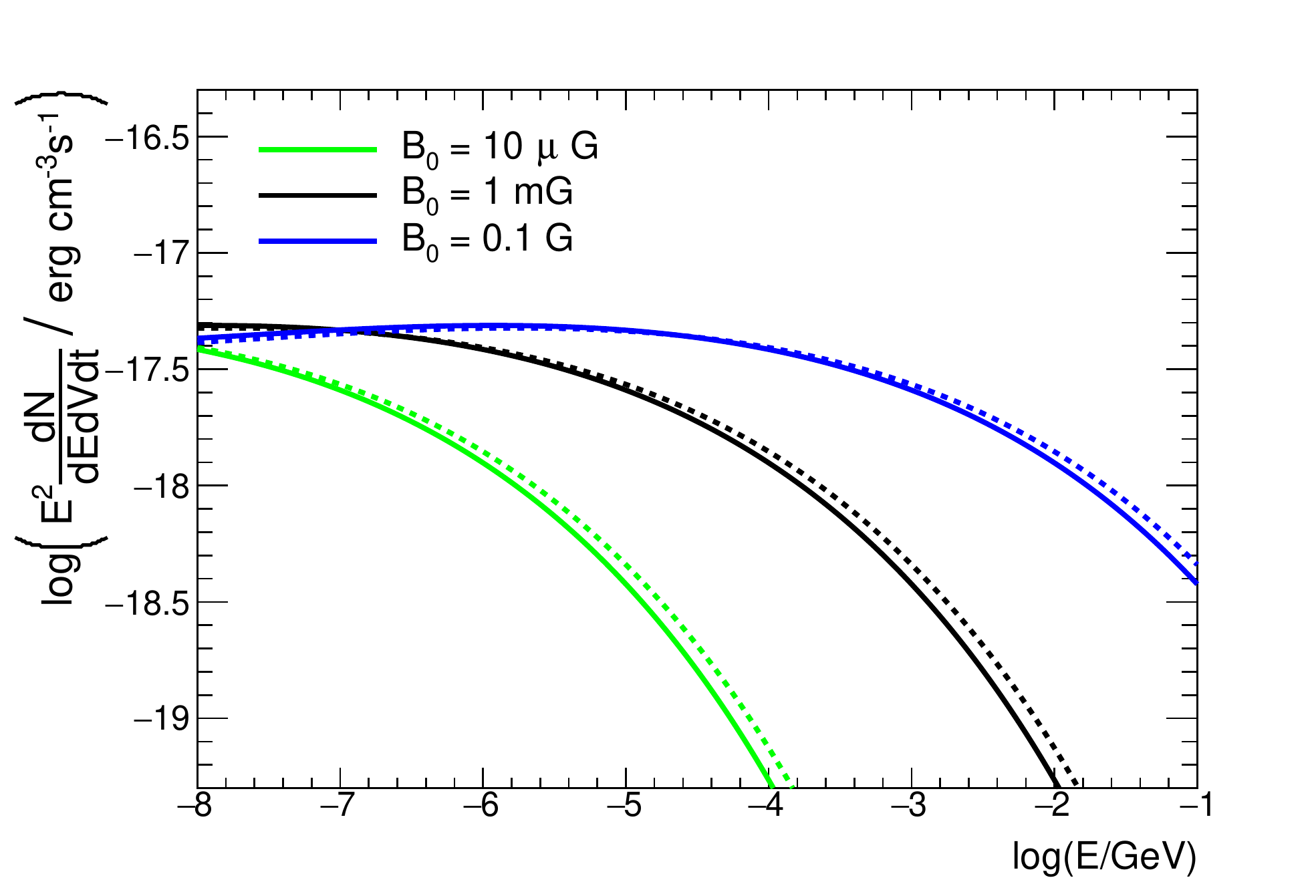}}
\caption{(a) Emissivities of gamma rays, muon neutrinos and electron neutrinos from $pp$ interactions. (b) Emissivities of synchrotron photons radiated by secondary electrons in uniform (solid) and turbulent (dashed) magnetic fields, for different field strengths, as indicated in the legend. In both panels, the spectrum of parent protons is given by Eq.~\eqref{eq:protonsS} with $\alpha_{\rm p}=2.0$, $\beta_{\rm p}=1.0$ and $E_{0,{\rm p}}=10^6$~GeV. Note that normalization to the energy density of parent protons and number density of ambient gas are the same as in Fig.~\ref{fig:spectra}.} 
\end{figure*}

\begin{figure*}
\centering
\subfigure[\label{fig:gammaNuB}]{\includegraphics[width=0.48\textwidth]{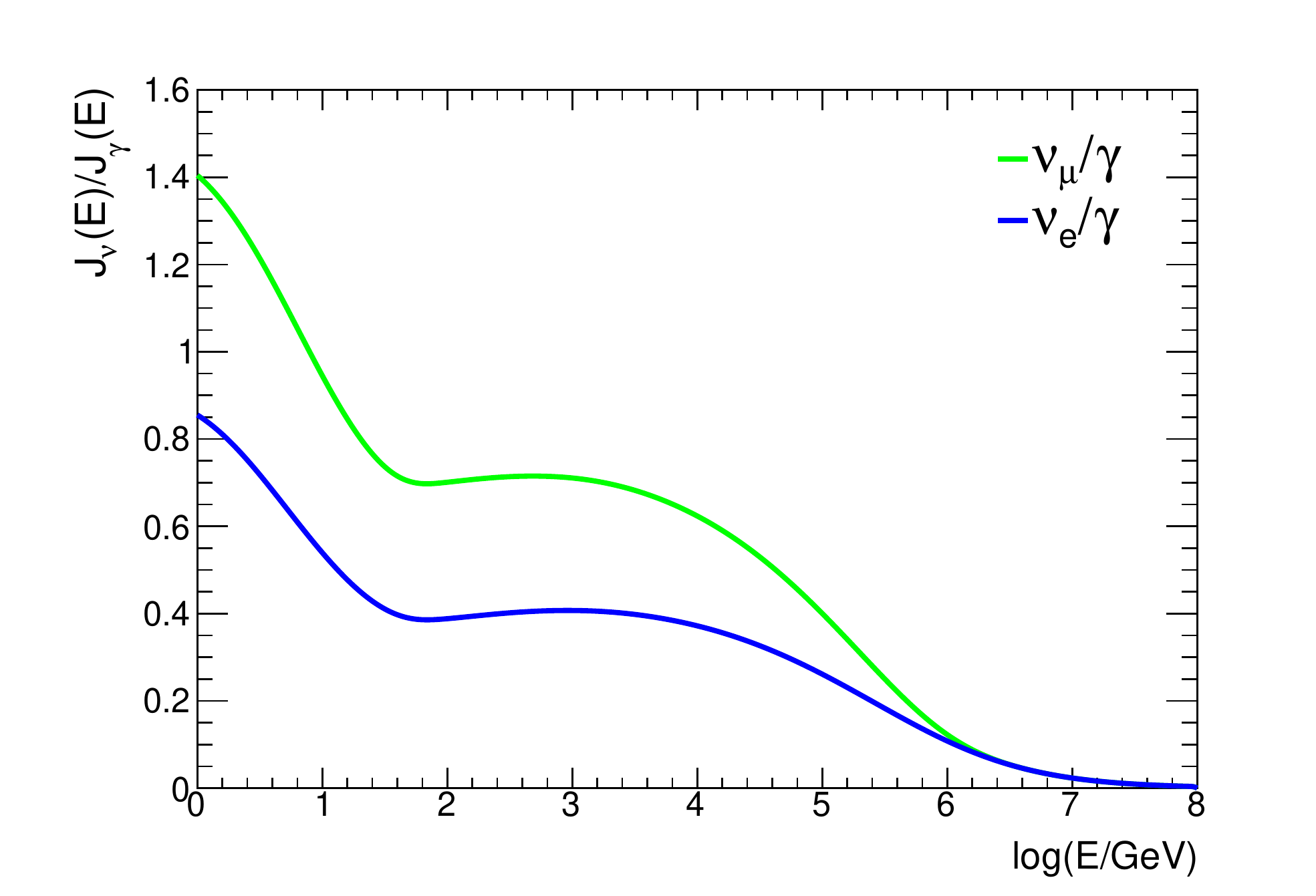}}
\caption{(a) Ratio among neutrino and gamma-ray spectra, as a function of energy. The spectrum of parent protons is given by Eq.~\eqref{eq:protonsS} with $\alpha_{\rm p}=2.0$, $\beta_{\rm p}=1.0$ and $E_{0,{\rm p}}=10^6$~GeV. Note that normalization to the energy density of parent protons and number density of ambient gas are the same as in Fig.~\ref{fig:spectra}.} 
\end{figure*}

\subsection{Spectral slopes}
\label{subsec:powerLaw}
We start with parametrizing the slopes $\alpha_{\rm s}$ of the energy distributions of secondary products in the form of Eq.~\eqref{eq:fitPLC}. In the case of pure power-law distribution of protons with slope $\alpha_{\rm p}$, because of the slight increase of the total cross-section of the inelastic $pp$ interactions, we expect somewhat harder gamma-ray and neutrino spectra \citep{kelner2006}. Namely, over a few decades in particle energy, the spectrum of gamma rays, neutrinos and electrons can be described with a power-law index $\alpha_{\rm s} \simeq \alpha_{\rm p}-0.1$. For the synchrotron radiation,
taking into account the radiative cooling of electrons, the photon index is $\alpha_{\rm sy} \simeq (0.5 \alpha_{\rm p}+0.95)$. 
In the case of the proton spectrum  described by Eq.~\eqref{eq:protonsS}, we investigate the connection among $\alpha_{\rm s}$ and $\alpha_{\rm p}$ by assuming a linear dependence between the two, namely:
\begin{equation}
\label{eq:fitLinear}
\alpha_{\rm s}=m\alpha_{\rm p}+q \ . 
\end{equation}
{Because of the fact that we obtain $\alpha_{\rm s}$ from a multi-frequency fitting procedure, the parameter $q$ shows a minor dependence on $\beta_{\rm p}$ and $E_{\rm 0,p}$. However, since at energies of secondaries much smaller than $E_{\rm 0,p}$, the parameter $\alpha_{\rm s}$ can be interpreted as the power-law index, the dependence should be very weak. Hence, for the broad-band spectra, we performed weighted average of the best-fit values with respect to $\beta_{\rm p}$ and $E_{0,{\rm p}}$, obtaining the following average fit parameters:  \\
\newline
(i) Gamma rays: 
\begin{equation}
\label{eq:ag}
\alpha_{\gamma}=0.94 \alpha_{\rm p}-0.15 \ .
\end{equation}

\noindent
(ii) Muon neutrinos: 
\begin{equation}
\label{eq:an}
\alpha_{\nu_\mu}=1.05 \alpha_{\rm p}-0.22 \ .
\end{equation}

\noindent
(iii) Electrons (and electron neutrinos): 
\begin{equation}
\label{eq:ae}
\alpha_{\rm e}=1.02 \alpha_{\rm p}-0.15 \ .
\end{equation}

\noindent
(iv) Synchrotron photons (radiated in the Gaussian turbulent field with $B_0$=1~mG):
\begin{equation}
\label{eq:ast}
\alpha_{\rm sy,t}=0.51 \alpha_{\rm p}+0.92 \ .
\end{equation}

Note that the above results are close but not identical with the results obtained for the pure power-law spectrum of protons without 
a break or a cut-off. This is explained by the systematics induced by fitting the multi-frequency spectra, and therefore facing  a degeneracy between the slope, the cut-off energy and the shape of the spectrum in the cut-off region.

\subsection{Spectral shapes in the cut-off region}
\label{subsec:bb}

To investigate the relations between the spectral shapes in the cut-off region, we model the relation between the parameter $\beta_{\rm p}$, which characterizes the spectrum of parent protons, and  $\beta_{\rm s}$, as obtained in the spectral fitting procedure. It is expected that the sharpening of the spectrum of protons in the cut-off region should be reflected, in one way or another, in the spectrum of secondaries. We  represent the link between these two parameters in the following form  
\begin{equation}
\label{eq:bsbp}
\beta_{\rm s}=\frac{\beta_{\rm p}}{a\beta_{\rm p}+b} \ ,
\end{equation}
where $a$ and $b$  are obtained through post-processing of the fitting results. The study performed over different combinations of $\alpha_{\rm p}$ and $E_{0,{\rm p}}$ shows significant dependence on $\alpha_{\rm p}$. This is demonstrated in Figs.~\ref{fig:scatterBetaA}-\ref{fig:scatterBetaC} where the $\beta_{\rm s} - \beta_{\rm p}$ relations are shown for fixed value of proton cut-off energy, i.e. $E_{0,{\rm p}}=10^5$~GeV. 
As expected, the dependence on $E_{0,{\rm p}}$ is rather weak: this can be seen in Fig.~\ref{fig:scatterBetaD} from the comparison of two curves corresponding to $E_{0,\rm p}=10^5$~GeV and $E_{0,{\rm p}}=10^8$~GeV (both are calculated for $\alpha_{\rm p}=2$). Therefore, for a practical purpose, below we present the $E_{\rm 0,p}$-independent parametrizations for  $\beta_{\rm s}$, which provide an accuracy better than 20\%, for any value of  $E_{0,\rm p}$ between $10^5$ and $10^{8}$~GeV. The resulting relations are: \\

\noindent
(i) Gamma rays: 
\begin{equation}
\label{eq:bg}
\beta_{\gamma}=\frac{\beta_{\rm p}}{(-0.7\alpha_{\rm p}+2.4)\beta_{\rm p}+0.1\alpha_{\rm p}+0.7}  \ .
\end{equation}

\noindent
(ii) Muon neutrinos: 
\begin{equation}
\label{eq:bn}
\beta_{\nu_\mu}=\frac{\beta_{\rm p}}{(-0.5\alpha_{\rm p}+2.1)\beta_{\rm p}+1.0} \ .
\end{equation}

\noindent
(iii) Electrons (and electron neutrinos): 
\begin{equation}
\label{eq:be}
\beta_{\rm e}=\frac{\beta_{\rm p}}{(-0.7\alpha_{\rm p}+2.4)\beta_{\rm p}+0.9}  \ .
\end{equation}

\noindent
(iv) Synchrotron photons:
\begin{equation}
\label{eq:bst}
\beta_{\rm sy,t}=\frac{\beta_{\rm p}}{(-1.5\alpha_{\rm p}+6.0)\beta_{\rm p}+2.0}  \ .
\end{equation}

\begin{figure*}
\centering
\subfigure[\label{fig:scatterBetaA}]{\includegraphics[width=0.48\textwidth]{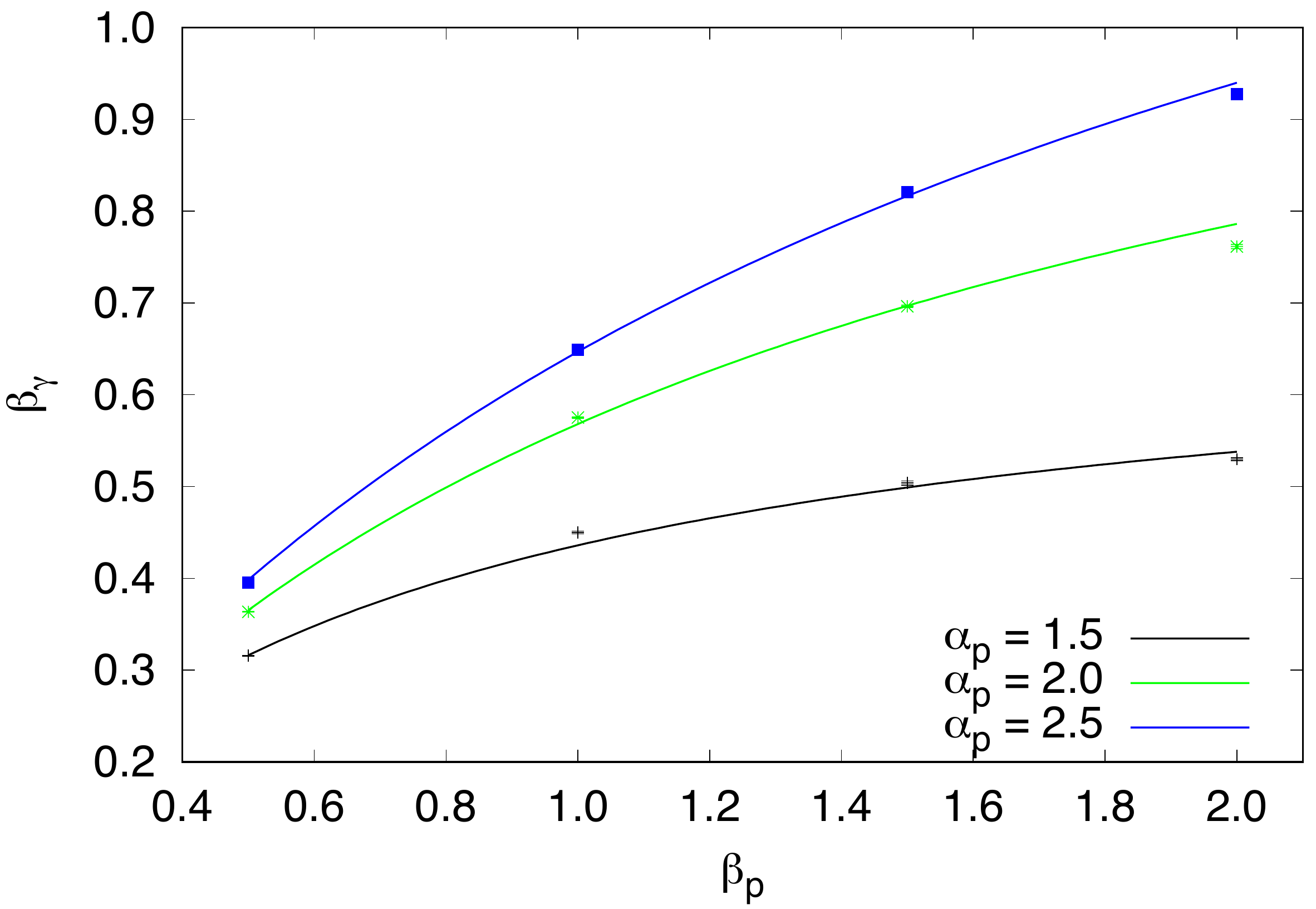}}
\subfigure[\label{fig:scatterBetaB}]{\includegraphics[width=0.48\textwidth]{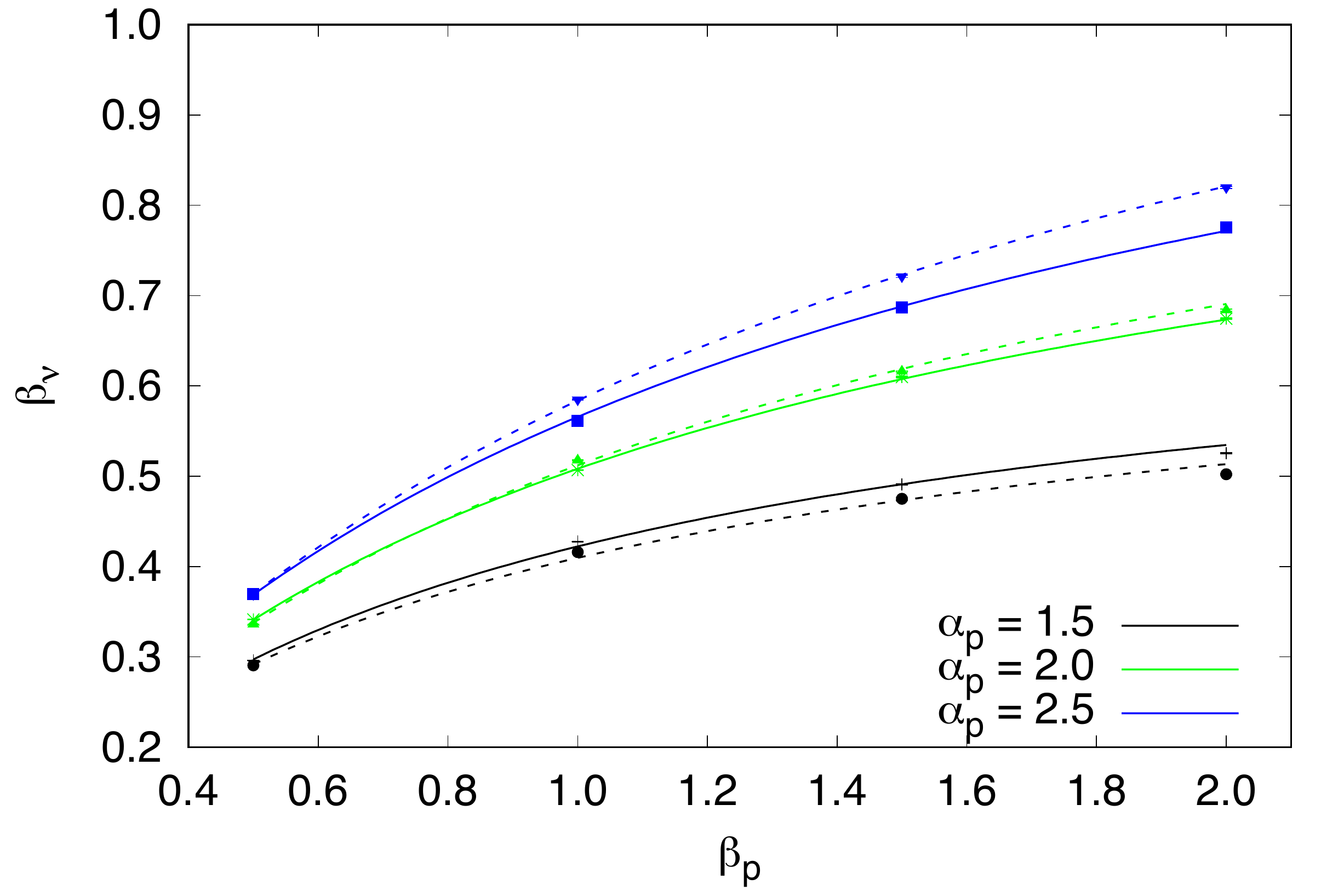}}
\subfigure[\label{fig:scatterBetaC}]{\includegraphics[width=0.48\textwidth]{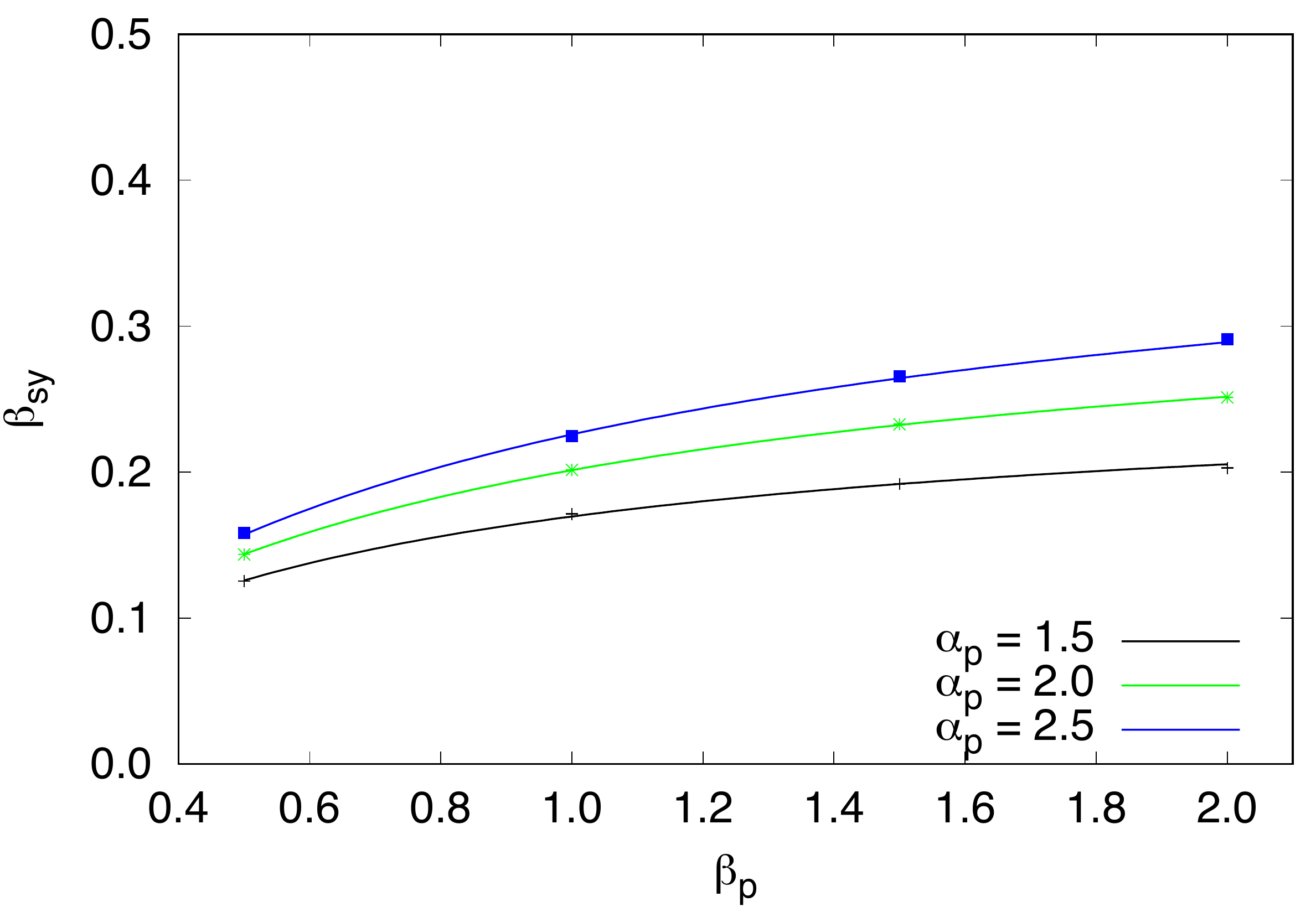}}
\subfigure[\label{fig:scatterBetaD}]{\includegraphics[width=0.48\textwidth]{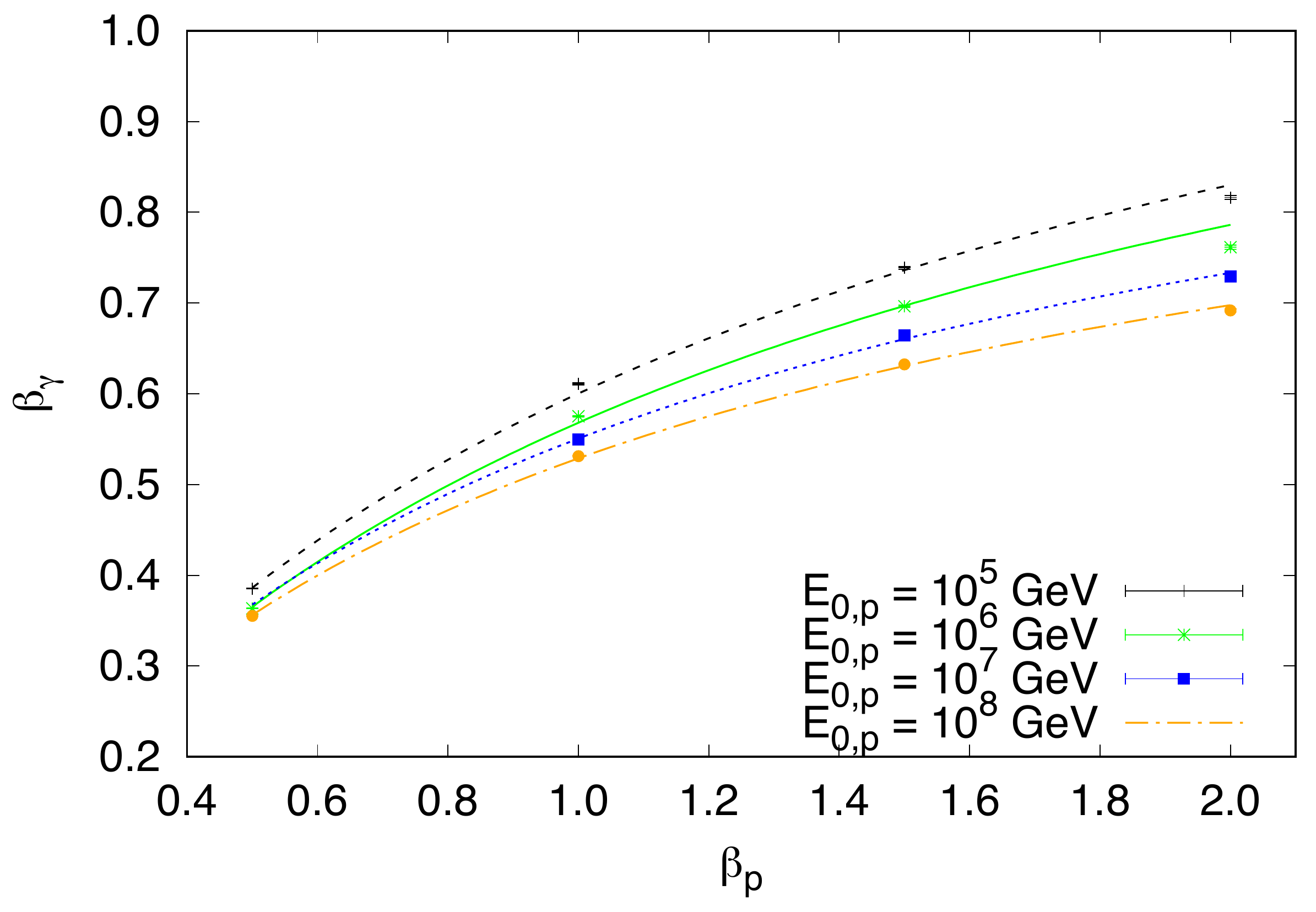}}
\caption{Correlation between values of $\beta_{\rm s}$, obtained in the fitting procedure of secondary particle spectra, and the values of $\beta_{\rm p}$ assumed for protons. Dots indicate simulated spectral values. The cut-off energy of protons is here fixed to $E_{0,{\rm p}}=10^6$~GeV. (a) Gamma rays; (b) Neutrinos (solid lines for $\nu_\mu$, dashed lines for $\nu_{\rm e}$); (c) Synchrotron photons (in turbulent magnetic field); (d) Gamma rays for different values of the proton cut-off energy and proton slope fixed to $\alpha_{\rm p}=2.0$. In each panel, lines refer to the analytical parametrization in the form of Eq.~\eqref{eq:bsbp}.}
\label{fig:betabeta}
\end{figure*}

\noindent
Note that for the given $E_{\rm p}$, the photons from the decay of $\pi^0$-mesons are produced with an average energy of  $\sim 0.1 E_{\rm p}$. The neutrinos from the decays of charged pions receive approximately twice less energy. However, at this point it is worth to stress that the exact amount of energy that each secondary receives depends on the spectral energy distributions of the parent protons, particularly in the cut-off region. In fact, at high energies, the $pp$ interactions proceeds in the multi-pion production regime, i.e. approximately half of the proton energy is given cumulatively to $\pi^+$, $\pi^-$ and $\pi^0$. Among these mesons, one is the so-called leading pion, namely it retains most of the energy that is available for the three. As a consequence, the decay products of the leading pion will exceed the average expectations, obtaining energies comparable to that of the pion itself. For this reason, the cut-off shape in neutrinos will differ from that in gamma rays, making accurate calculations necessary to describe how energy gets shared in the interaction process. 

\subsection{Connecting cut-off energies}
\label{subsec:be}
Since in the cut-off region, the spectral shape of the secondaries does not mimic exactly the shape of the proton spectrum, i.e. $\beta_{\rm s}$ and $\beta_{\rm p}$ differ, the relation between $E_{0,\rm s}$ and $E_{0,\rm p}$ is not linear but depends on $\beta_{\rm s}$. A rather weak dependence on $E_{\rm 0, p}$ also is expected, but, as in the previous section, for a practical purpose we ignore this slight effect and provide simple relations which with a good, better than 20\%, accuracy can be applied to a broad range of $E_{\rm 0, p}$ between $10^5$ to $10^7$~GeV. In turn, we observe a non negligible dependency on $\alpha_{\rm p}$, as shown in Fig.~\ref{fig:betaEcut}, that leads us to the following relations: \\

\noindent
(i) Gamma rays: 
\begin{equation}
\label{eq:esbg}
\log_{10} \left( \frac{E_{0,\gamma}}{E_{0,{\rm p}}} \right) = (-1.15\alpha_{\rm p}+3.30) \beta_{\gamma}+(1.33\alpha_{\rm p}-4.61) \ .
\end{equation}

\noindent
(ii) Muon neutrinos: 
\begin{equation}
\label{eq:esbn}
\log_{10} \left( \frac{E_{0,\nu_\mu}}{E_{0,{\rm p}}} \right) =(-1.29\alpha_{\rm p}+3.90) \beta_{\nu_\mu}+(1.31\alpha_{\rm p}-5.05) \ .
\end{equation}

\noindent
(iii) Electrons (and electron neutrinos): 
\begin{equation}
\label{eq:esbe}
\log_{10} \left( \frac{E_{0,{\rm e}}}{E_{0,{\rm p}}} \right) =(-1.48\alpha_{\rm p}+4.34) \beta_{\rm e}+(1.50\alpha_{\rm p}-5.43) \ .
\end{equation}

\noindent
(iv) Synchrotron photons:
\begin{equation}
\label{eq:esbst}
\log_{10} \left( \frac{E_{0,{\rm sy,t}}}{E_{0,{\rm p}}} \right) = (-11.36\alpha_{\rm p}+34.58) \beta_{\rm sy,t}+(4.09\alpha_{\rm p}-21.72) \ .
\end{equation}

Note that the parameters $\beta_\gamma$, $\beta_\nu$, $\beta_{\rm e}$, and $\beta_{\rm sy}$ entering in the above equations can be found from Eqs.~\eqref{eq:be}-\eqref{eq:bn}. Thus, the relation between $E_{\rm 0, s}$ and $E_{\rm 0, p}$ are determined only by the parameters $\alpha_{\rm p}$ and $\beta_{\rm p}$ of parent protons. 

\begin{figure*}
\centering
\subfigure[]{\includegraphics[width=0.48\textwidth]{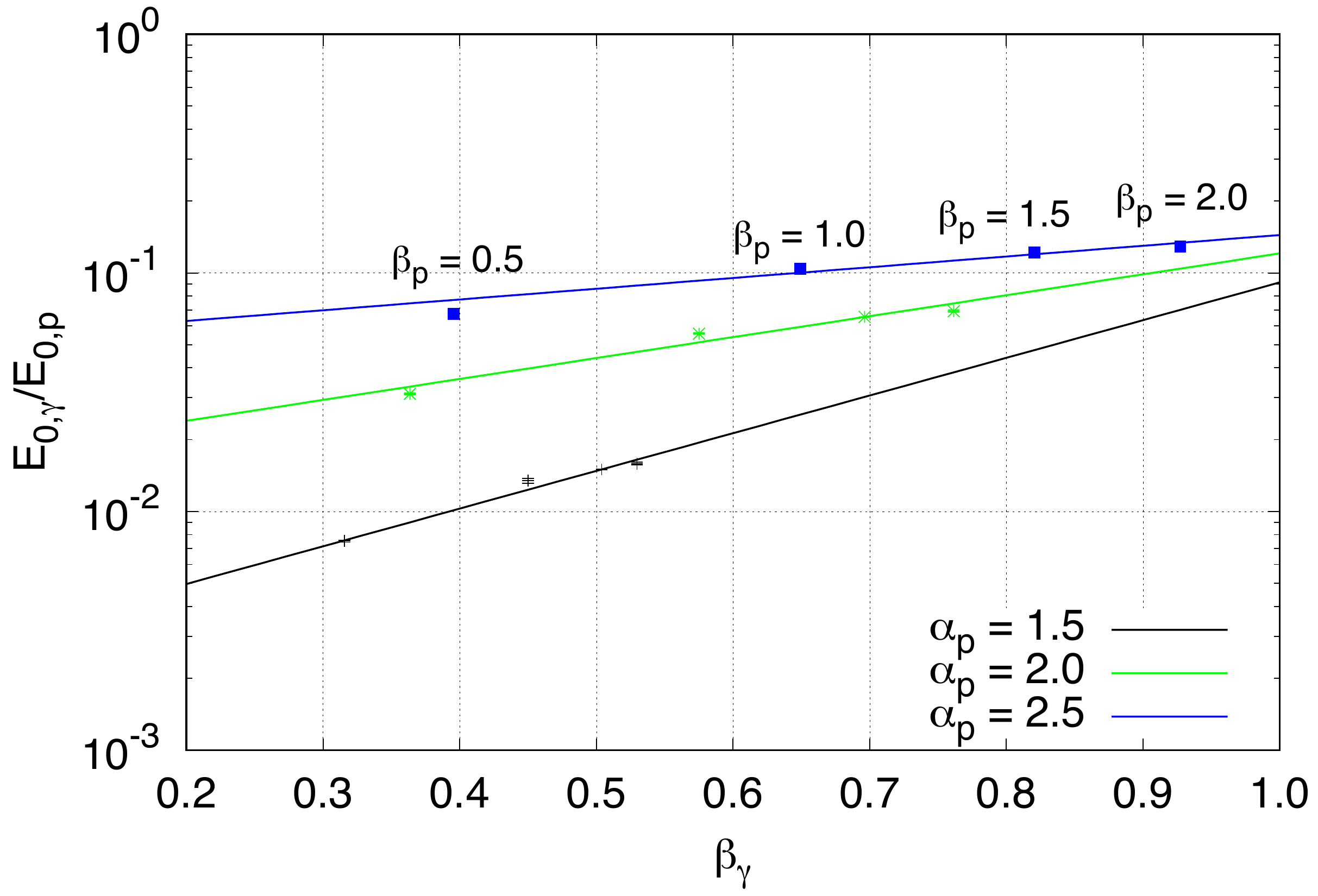}}
\subfigure[]{\includegraphics[width=0.48\textwidth]{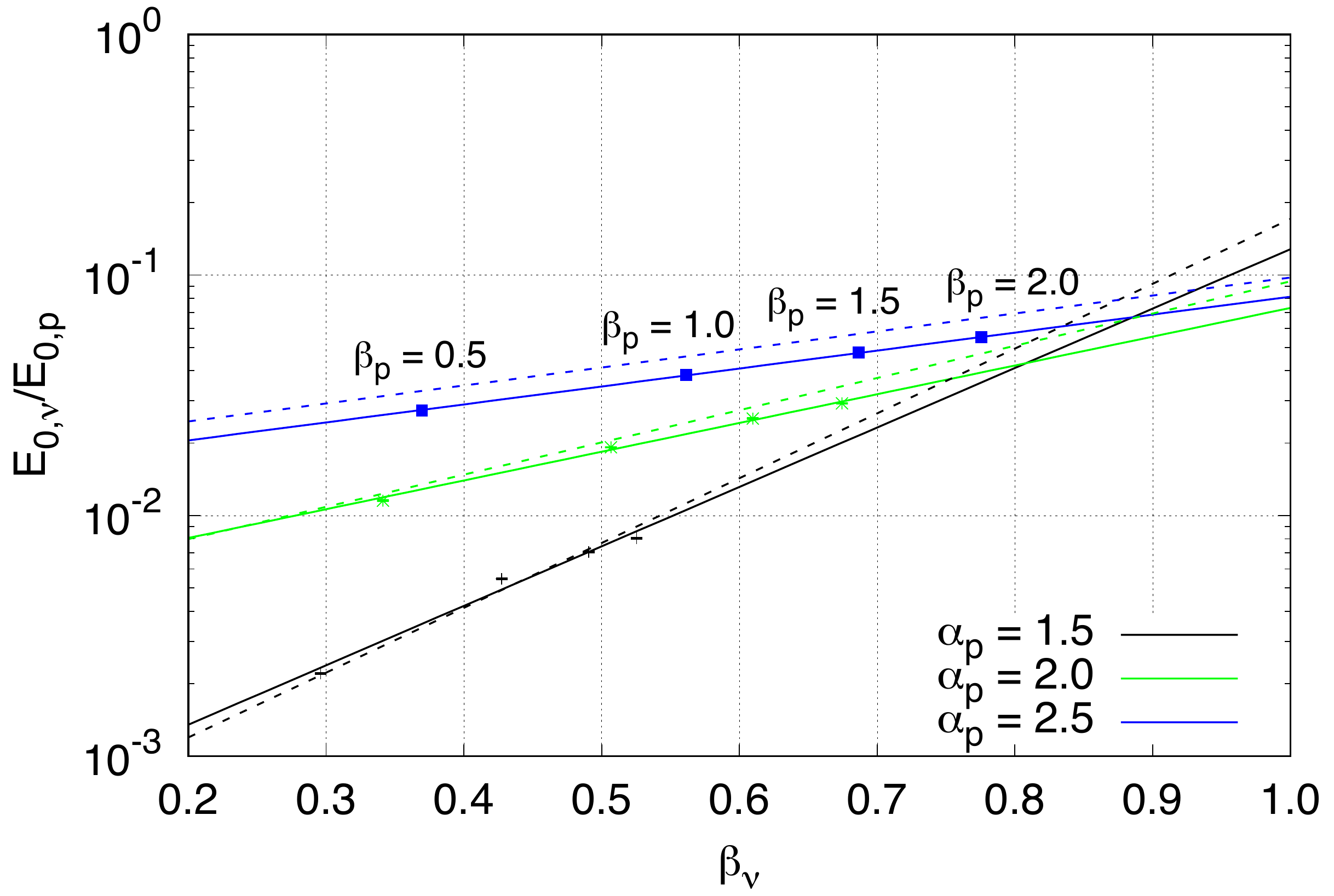}}
\subfigure[]{\includegraphics[width=0.48\textwidth]{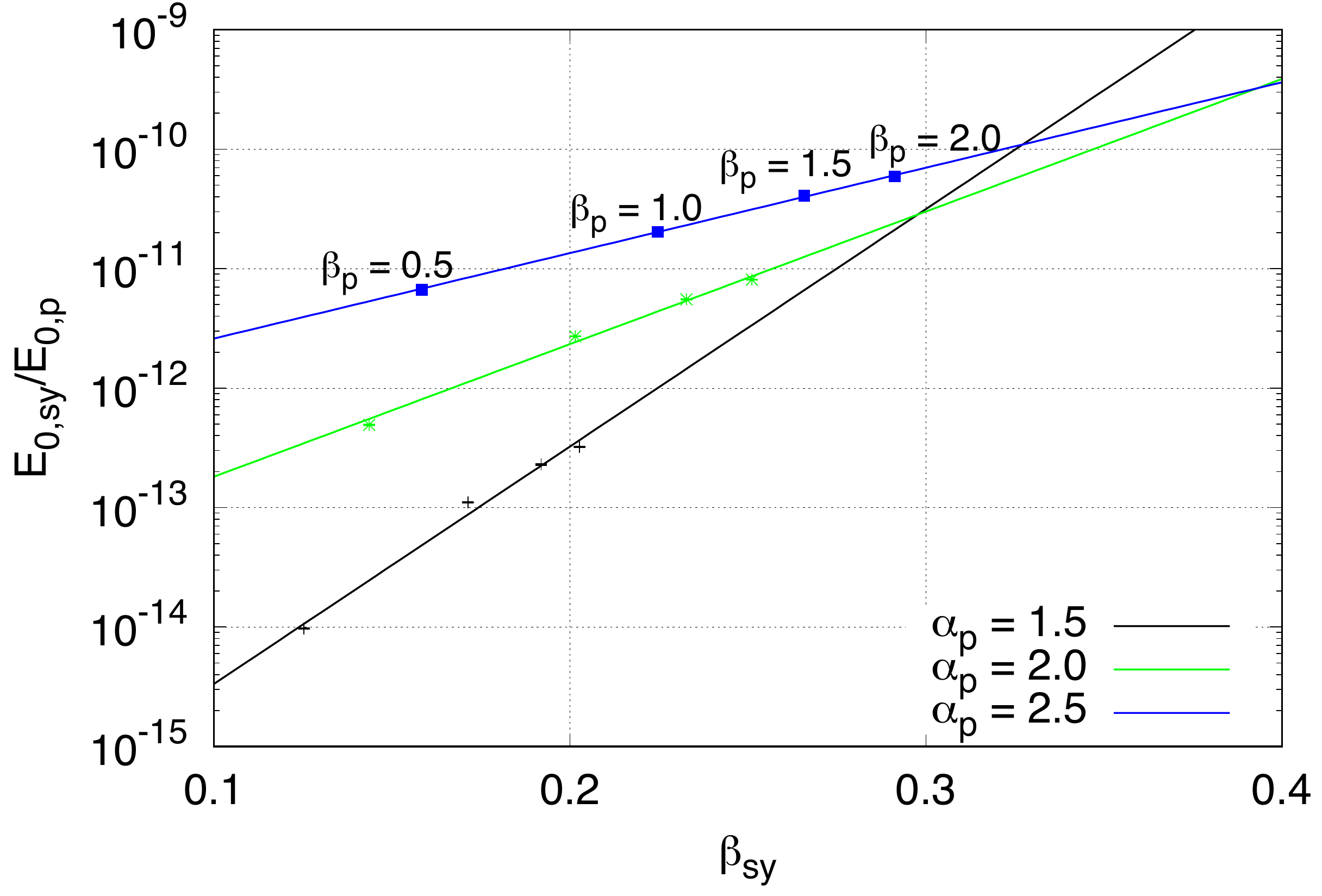}}
\caption{Ratios $E_{0,{\rm s}}/E_{0,{\rm p}}$ as a function of $\beta_{\rm s}$ calculated for $E_{\rm 0, p}=10^6$~GeV and three different values of $\alpha_{\rm p}$=1.5, 2.0, 2.5. (a) Gamma rays (dashed line corresponds to $E_{\rm 0, p}=10^7$~GeV; (b) Neutrinos (solid lines for $\nu_\mu$, dashed lines for $\nu_{\rm e}$); (c) Synchrotron photons (for $B_0$=1 mG).  Individual dots indicated at each curve refer to different values of $\beta_{\rm p}$=0.5, 1.0, 1.5, 2.0.}
\label{fig:betaEcut}
\end{figure*}


\section{Narrow proton distributions}
\label{sec:maxwell}
In previous sections,  the energy spectra of the secondary particles have been studied for a rather broad variety of the spectra of  parent protons, including hard energy distributions with $\alpha_{\rm p}=1.5$. In certain astrophysical environments, physical conditions that would lead to even harder acceleration spectra of protons could possibly be realized.
In particular, in the case of colliding stellar winds, under certain conditions the spectral slope might be as hard as $\alpha_{\rm p}=1$ (\citet{bykov2013}, but see also \citet{vieu2020}). Moreover, in some acceleration scenarios, e.g. at the magnetic reconnection \citep{lazarian2015}, a very narrow distribution of particles can be formed with negative values of  $\alpha_{\rm p}$. In an extreme case of $\alpha_{\rm p}=-2$, such a spectrum resembles the relativistic Maxwellian distribution. Note that this is a formal definition of the functional form of particle distribution and should not be misinterpreted as distribution achieved as a result of thermal equilibrium.  

In this Section, we will consider primary protons narrowly distributed in energy, i.e. according to a Maxwell-J\"uttner distribution \citep{juttner1911}. For protons at a characteristic thermal scale $kT$, $k$ being the Boltzmann constant and $T$ being the system temperature, we define $\theta=kT/(m_{\rm p}c^2)$ ($m_{\rm p}$ being the proton's mass), and the modified Bessel function of the second kind $K_2(1/\theta)$, such that the energy distribution of protons reads as
\begin{equation}
\label{eq:maxwell}
\frac{dN_{\rm p}}{dE_{\rm p}} =  \left( \frac{E_{\rm p}}{kT m_{\rm p}c^2} \right) \frac{\exp \left[-\frac{E_{\rm p}}{kT} \right]}{K_2(1/\theta)} \sqrt{\left(\frac{E_{\rm p}}{m_{\rm p}c^2} \right)^2 -1}   \ .
\end{equation}
Formally, this distribution applies to protons at thermal equilibrium. In our case, however, we will consider such a distribution as representative of an acceleration scenario with injection slope much harder than $2$. In fact, for a gas of relativistic particles, Eq.~\eqref{eq:maxwell} scales as $dN_{\rm p}/dE_{\rm p} \propto E_{\rm p}^2 \exp \left[-(E_{\rm p}/kT) \right]$. On the other hand, in the non-relativistic case, the classical Maxwell distribution is recovered. Note that, with respect to the exponentially suppressed power law of Sec.~\ref{sec:param}, a narrow distribution is a limiting case of Eq.~\eqref{eq:protonsS} for $\beta_{\rm p} \gg 1$. \\
Following the methods outlined in Sec.~\ref{sec:methods2}, we obtain the spectral distributions of secondary particles resulting from $pp$ collisions, as shown in Fig.~\ref{fig:maxwell}. In order to compare with the results shown previously for cut-off power-law energy distributions of protons, we adopt the same normalization used earlier, i.e. we fix the energy density of protons above 100~GeV to $w_{\rm p}=1$~erg~cm$^{-3}$.
In Fig.~\ref{fig:maxwell} we show the resulting broadband spectral energy distribution of gamma rays, neutrinos as well as the secondary synchrotron photons, compared to those produced by an exponential cut-off distribution of protons, defined by the following parameters:  $\alpha_{\rm p}=2$, $\beta_{\rm p}=1$ and $E_{\rm 0,p}=10^5$~GeV. In both cases, the synchrotron radiation is calculated for the Gaussian turbulent magnetic field of strength $B_0=1$~mG. As expected, a clear difference emerges among the two cases, namely the energy extent of the energy distribution of secondaries, which is narrower in the Maxwellian case by about two orders of magnitude. \\

\begin{figure*}
\centering
\subfigure[]{\includegraphics[width=0.497\textwidth]{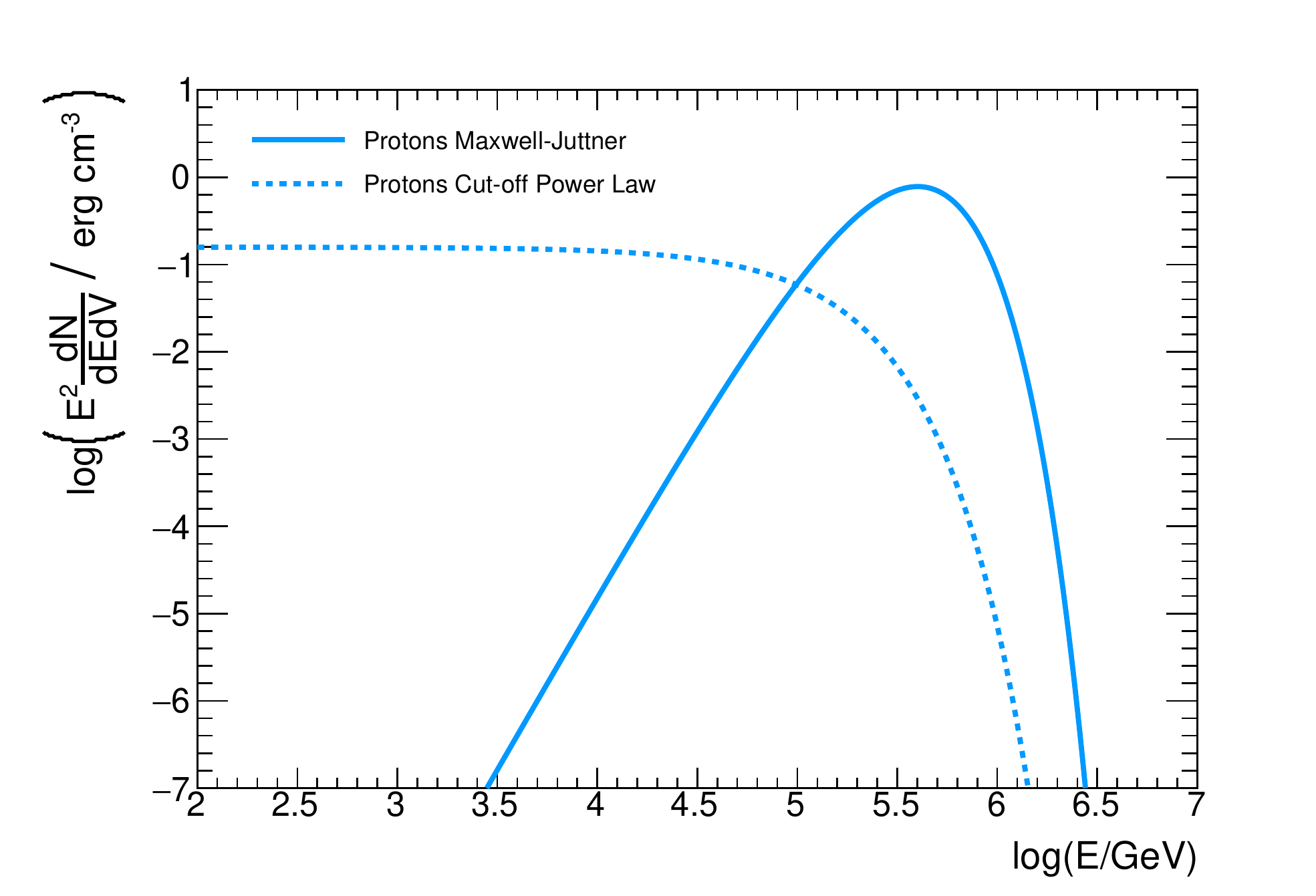}}
\subfigure[]{\includegraphics[width=0.497\textwidth]{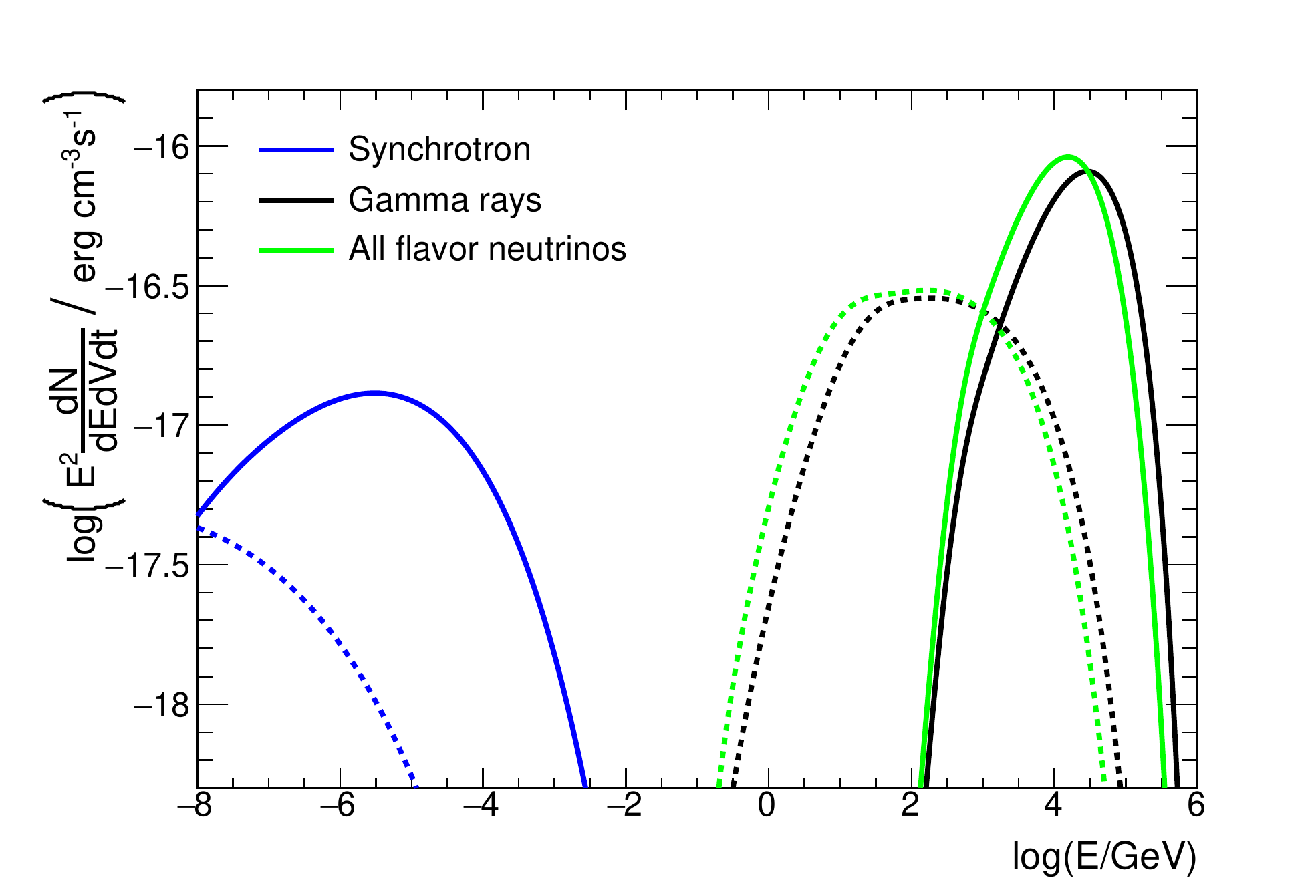}}
\caption{Spectral energy distribution of secondaries resulting from $pp$ collisions, in the hypothesis of primary protons following a Maxwellian-like distribution (see Eq.~\eqref{eq:maxwell}) with $kT_{\rm p}=10^5$~GeV (solid lines), or a non-thermal cut-off power law distribution (see Eq.~\eqref{eq:protonsS}) with $\alpha_{\rm p}=2$, $\beta_{\rm p}=1$, and $E_{0,{\rm p}}=10^5$~GeV (dashed lines). (a) Protons; (b) Synchrotron photons ($B_0=1$~mG), gamma rays and all flavor neutrinos.}
\label{fig:maxwell}
\end{figure*}

\section{Summary and discussion}
\label{sec:disc}
The lack of observational evidence for the CR sources responsible for particles detected on Earth with energy between $10^{15}$ and $10^{17}$~eV is one of the major open problems in astroparticle physics today. For this reason, it appears extremely timely to investigate the physical processes that these particles undergo, among which $pp$ collisions are a guaranteed interaction process, given the abundance of free protons in the interstellar medium. In this paper, we performed a spectroscopic analysis of the secondary particles produced in $pp$ collisions, with a special focus on the cut-off region, with the aim of providing useful and easy-to-use parametrizations to the observational community. For this study, we adopted the latest measurements of the interaction cross-section, and we considered a uniform density of the proton target. The latter assumption allows a direct rescaling of secondary spectra shown here with the proton density of interest. \\
We started by considering an energy distribution of the primary protons in the form of an exponentially suppressed power law. We allowed the proton slope, cut-off energy and shape to vary, and we investigated the effects of such variations on the spectra of secondaries.
Thanks to a multi-frequency fitting procedure, we have provided simple analytical parametrizations, describing the connection among: i) $\alpha_{\rm s}$ and $\alpha_{\rm p}$, ii) $\beta_{\rm s}$ and $\beta_{\rm p}$, and iii) $E_{0, \rm s}$ and $E_{0, \rm p}$. This set of relations can be used to infer the spectrum of primary protons from the spectrum of radiation, specified through $\alpha_{\rm s}$, $\beta_{\rm s}$ and $E_{0, \rm s}$. \\
As a result, we achieved the following conclusions: \\
(i) With respect to the shape of the cut-off, we observe that the cut-off of synchrotron photons radiated by secondary electrons is shallower than any other secondary particles produced in the interaction. In particular, if electrons lose energy in turbulent magnetic fields, the cut-off shape of synchrotron photons is even milder than in the case of uniform magnetic field. This condition makes the emitted radiation particularly interesting for the exploration of cosmic accelerators, since a sharp cut-off in protons is observed as a mild decrease in synchrotron photons. It was shown that PeV protons in mG magnetic field are able to produce radiation at few keV. Clearly, to safely identify a proton accelerator, one should be able to exclude the leptonic origin of the radiation: this might be feasible in passive molecular clouds, namely those clouds located far enough from the accelerator, that the highest energy primary electrons would be prevented from getting there due to the severe energy losses they undergo, while primary protons would. We suggest that the most efficient strategy to look for PeV and multi-PeV accelerators is through hard X rays from dense molecular clouds illuminated by a distant accelerator. \\
(ii) With respect to the energy of the cut-off, we confirm kinematic arguments that predict for gamma rays a cut-off energy of about a tenth of the proton cut-off energy, while for neutrinos and secondary electrons about a twentieth. However, we observe that the cut-off energy inferred from the spectra of secondary particles depends on the spectral shape of the primary protons. Such an effect requires detailed calculations for the spectra of secondaries. The parametrizations given here allows to avoid performing extensive integrations, while providing an accurate description of the relations among parent particles and the different emerging species, in terms of both spectral shape and energy transfer. For practical purposes, these formulas can be used from secondaries to primaries, and are hence crucial for inferring the physical processes ongoing at the source, including acceleration and propagation. \\
(iii) With respect to the spectral slope, we found results in line with theoretical expectations in the pure power-law region, i.e. the slope of secondary electrons, neutrinos and gamma rays is harder than that of primary protons by $\sim 0.1$. We observe a contained systematics induced by the multi-frequency modeling. \\
We also considered the situation where the energy of primary protons rather follows a \textquoteleft\textquoteleft Maxwellian-type\textquoteright\textquoteright $\,$ distribution, namely it is narrowly peaked in energy. We showed that this situation can mimic the effects induced by a proton energy distribution in the form of a power law with an exponential suppression, though generally resulting in narrower energy distributions. 

\appendix
\section{Kernel functions for secondary particles}
\label{sec:app}
In this Appendix, we report the kernel functions that were adopted to derive the spectra of secondary particles produced in $pp$ collisions, as derived by \citet{kelner2006}. In the following, we fix $L=\ln(E_{\rm p}/{\rm TeV})$, $E_{\rm p}$ being the energy of the primary proton. To compute the amount of gamma rays per collision, we define $x=E_{\gamma}/E_{\rm p}$, $E_{\gamma}$ being the energy of the emerging gamma ray. Then, the number of gamma rays in the interval (x,x+dx) is given by:
\begin{equation}
F_\gamma(x,E_{\rm p})= B_\gamma \frac{\ln(x)}{x}  \left( \frac{1-x^{\beta_\gamma}}{1+k_\gamma x^{\beta_\gamma} (1-x^{\beta_\gamma})} \right)^4 
\left[ \frac{1}{\ln(x)} - \frac{4 \beta_\gamma x^{\beta_\gamma}}{1-x^{\beta_\gamma}} - \frac{4 k_\gamma \beta_\gamma x^{\beta_\gamma} (1-2x^{\beta_\gamma})}{1+k_\gamma x^{\beta_\gamma} (1-x^{\beta_\gamma})}  \right] \ ,
\end{equation}
where
\begin{equation}
B_\gamma=1.30+0.14 L +0.011 L^2 \ ,
\end{equation}
\begin{equation}
\beta_\gamma=(1.79+0.11 L + 0.008 L^2)^{-1} \ ,
\end{equation}
\begin{equation}
k_\gamma=(0.801+0.049 L + 0.014 L^2)^{-1} \ .
\end{equation}
These expressions were adopted in Eq.~\eqref{eq:phigamma}.\\
On the other hand, setting $x=E_{\nu_\mu}/E_{\rm p}$ and $y=x/0.427$, the number of muon neutrinos in the interval (x,x+dx) per collision, emerging from the direct pion decay, can be computed through:
\begin{equation}
F_{\nu_\mu}^{(1)}(x,E_{\rm p})= B^\prime \frac{\ln(y)}{y}  \left( \frac{1-y^{\beta^\prime}}{1+k^\prime y^{\beta^\prime} (1-y^{\beta^\prime})} \right)^4 
\left[ \frac{1}{\ln(y)} - \frac{4 \beta^\prime y^{\beta^\prime}}{1-y^{\beta^\prime}} - \frac{4 k^\prime \beta^\prime y^{\beta^\prime} (1-2y^{\beta^\prime})}{1+k^\prime y^{\beta^\prime} (1-y^{\beta^\prime})}  \right] \ ,
\end{equation}
where
\begin{equation}
B^\prime=1.75+0.204 L +0.010 L^2 \ ,
\end{equation}
\begin{equation}
\beta^\prime=(1.67+0.111 L + 0.0038 L^2)^{-1} \ ,
\end{equation}
\begin{equation}
k^\prime=1.07-0.086 L + 0.002 L^2 \ .
\end{equation}
Lastly, defining $x=E_{\rm e}/E_{\rm p}$, the number of electrons produced in the interval (x,x+dx) from the muon decay is given by
\begin{equation}
F_{\rm e}(x,E_{\rm p})= B_{\rm e} \frac{(1+k_{\rm e} (\ln x)^2)^3}{x(1+0.3/x^{\beta_{\rm e}})} (-\ln(x))^5 \ ,
\end{equation}
where
\begin{equation}
B_{\rm e}=(69.5+2.65 L +0.3 L^2)^{-1} \ ,
\end{equation}
\begin{equation}
\beta_{\rm e}=(0.201+0.062 L + 0.00042 L^2)^{-1/4} \ ,
\end{equation}
\begin{equation}
k_{\rm e}=\frac{0.279+0.141 L + 0.0172 L^2}{0.3+(2.3+L)^2} \ .
\end{equation}
Note that the spectrum of muon neutrinos from the decay of muons, $F_{\nu_\mu}^{(2)}(x,E_{\rm p})$, is described by the same function, with $x=E_{\nu_\mu}/E_{\rm p}$. These expressions were adopted in Eq.~\eqref{eq:phinu}.

\acknowledgments

SC acknowledges the support of the fellowship \textquoteleft\textquoteleft L'Or\'eal Italia Per le Donne e la Scienza\textquoteright\textquoteright. SG acknowledges support from Agence Nationale de la Recherche (grant ANR- 17-CE31-0014) and from the Observatory of Paris (Action F\'ed\'eratrice CTA).

\bibliography{biblio}{}
\bibliographystyle{aasjournal}



\end{document}